\documentclass[a4,10pt]{article}
\usepackage{amsmath}
\usepackage{amssymb}
\usepackage{graphicx,epstopdf}
\usepackage{subfig}
\usepackage{caption}
\usepackage{multirow}
\providecommand{\keywords}[1]{\textit{Keywords---} #1}

\def\qslash{q\!\!\!\slash }
\def\xslash{x\!\!\!\slash }

\begin{document}

\title{Axial vector transition form factors of $ N \rightarrow \Delta $  in QCD}

\author{
A. Kucukarslan$^1$ \thanks{ e-mail: akucukarslan@comu.edu.tr},
 U. Ozdem$^1$ \thanks {e-mail: ulasozdemm@gmail.com,(Corresponding author)},
A. Ozpineci$^2$  \thanks {e-mail: ozpineci@metu.edu.tr} \\
\\
  \small$^1$ Physics Department, Canakkale  Onsekiz Mart University, 17100 Canakkale, Turkey\\
 \small $^2$ Physics Department, Middle East Technical University, 06531 Ankara, Turkey}

 \date{\today}

\maketitle
\begin{abstract}
The isovector axial vector form factors of $ N \rightarrow  \Delta $ transition are calculated by employing Light-cone QCD sum rules. The analytical results are analysed by both the conventional method, and also by a Monte Carlo based approach which allows one to scan all of the parameter space.
The predictions are also compared with the results in the literature, where available.
Although the Monte Carlo analysis predicts large uncertainties in the predicted results, the predictions obtained by the conventional analysis
are in good agreement with other results in the literature.
\end{abstract}

\section{Introduction}

Form factors are important properties of hadrons that contain information about their dynamics and internal structure. In this work, isovector axial form factors for the $N \rightarrow \Delta$ transition
are studied. These form factors are also important in the pion production off the nucleon, where the $\Delta$
baryon appears as an intermediate resonance, and can strongly influence the production rates
(see e.g. \cite{Amaro:2008hd}). The N to $\Delta $ transition has the advantage that the $\Delta$
is the dominant, clearly accessible nucleon resonance \cite{Alexandrou:2006mc}. The information contained in the axial isovector N to $\Delta $ transition form factors can be considered as supplementary  to that from electromagnetic  form factors. It is also possible to check the validity of the off-diagonal Goldberger-Treiman relation using a measurement of the isovector axial vector form factors for N to $\Delta $ transition \cite{Aliev:2007pi, Hernandez:2010bx}.

The first experimental results of $ N \rightarrow \Delta $ axial vector form factors comes from $G^0$ collaborations at JLAB \cite {Androic:2012doa} based on parity-violating electron scattering. At  $Q^2 $ = 0.34 $GeV^2$, the obtained  value of axial form factors determined from the hydrogen asymmetry in inelastic electron-proton scattering is  $G_A^{N\Delta}$ = - $ 0.05\pm(0.35)_{stat} \pm(0.34)_{sys} \pm(0.06)_{th}$. 
Furthermore, several theoretical investigations of $ N \rightarrow \Delta $ axial vector form factors are carried out, on lattice QCD \cite{Alexandrou:2006mc},
in quark models \cite{BarquillaCano:2007yk}, using light-cone
QCD sum rules \cite{Aliev:2007pi},
chiral perturbation theory ($\chi$PT) \cite{Geng:2008bm, Procura:2008ze}
or weak single pion production \cite{Hernandez:2010bx, Amaro:2008hd,Graczyk:2009qm}.

To study form factors, a non-perturbative method is necessary. The light cone QCD sum rules (LCSR) is
one of these non-perturbative methods that has been successfully applied to study non-perturbative hadronic properties   \cite{Chernyak:1990ag, Braun:1988qv, Balitsky:1989ry}.  In LCSR, the properties of the hadron under study are expressed in terms of the properties of the vacuum and the light cone distribution amplitudes(DAs) of the hadron.

Since the form factors are expresses in terms of the properties of the vacuum and DAs of the hadron, any uncertainty in these parameters should be reflected in the obtained predictions, on top of any other uncertainties inherent in the  sum rules approach. A
 Monte Carlo based approached is proposed   to
estimate the errors due to the uncertainties of these parameters by scanning the parameter space in \cite{Leinweber:1995fn}. In this work,  LCSR will be used to study the $N \rightarrow \Delta$ transition and
the Monte Carlo based method will be applied to estimate the errors in the form factors by extending the method proposed in \cite{Leinweber:1995fn} to the study of form factors.

The paper is organized as follows. In section II, the calculation of the sum rules in LCSR will be presented.
In section III, Monte Carlo analysis to estimate the uncertainties in the predictions
due to the input parameters will be presented. In the last section, we conclude our work with a discussion of our results.

 \section{Formulation of baryon axial form factors}
The LCSR for axial form factors of $N \rightarrow \Delta$ transition is derived from the following vacuum to nucleon correlation function:
\begin{eqnarray}\label{corrf}
	\Pi_{\mu\nu}(p,q)=i\int d^4 x e^{iqx} \langle 0 |T[J_{\mu}^{\Delta}(0)A_\nu^{3}(x)]|N(p,s)\rangle,
\end{eqnarray}
 where $J_{\mu}^\Delta(x)$ is an interpolating current for the $\Delta$, and $A^3_\nu$ is an axial vector-isovector current defined as
\begin{eqnarray}
A^3_\nu(x) = \frac12 \bigg( \bar u(x) \gamma_\nu \gamma_5 u(x) -  \bar d(x) \gamma_\nu \gamma_5 d(x) \bigg)
\end{eqnarray}

The correlation function given in Eq. (\ref{corrf}) can be calculated  in terms of hadronic properties, so called the hadronic representation, if $p^2>0$ and $(p+q)^2>0$,
and also in terms of QCD parameters and several distribution amplitudes (DAs) of the $N$ baryon, so called the QCD representation, in the
deep Euclidean region $p^2\rightarrow - \infty$ and $(p+q)^2 \rightarrow - \infty$.

The hadronic representation of the correlation function can be obtained by
inserting a complete set of states with the same quantum numbers as the
interpolating currents.
\begin{eqnarray}\label{fiz1}
   \Pi_{\mu\nu}(p,q) = \sum_{s'}\frac{\langle 0 |j_{\mu}^{\Delta}|\Delta(p',s')\rangle \langle \Delta(p',s')| A_\nu^3|N(p,s)\rangle }{M_{\Delta}^2-p'^2} +...
  \end{eqnarray}
where $M_\Delta$ is the mass of the $\Delta$ baryon.
The $N \to \Delta$ matrix element appearing in Eq. (\ref{fiz1}) can be written in terms of four invariant form factors as follows  \cite{Adler:1968tw,Adler:1975mt, LlewellynSmith1972261};
\begin{eqnarray}
\langle \Delta(p',s')| A_\nu^{3}| N(p,s)\rangle &= i \overline{\upsilon}^{\lambda}(p',s')\bigg[\left\{\frac{C_3^{N \Delta}(q^2)}{M_N}\gamma_\mu
+ \frac{C_4^{N \Delta}(q^2)}{M_N^{2}} p'_\mu\right\}
(g_{\lambda\nu} g_{\rho\mu} - g_{\lambda\rho} g_{\mu\nu})q^\rho\nonumber \\
&+ C_5^{N \Delta}(q^2) g_{\lambda\nu}+\frac{C_6^{N \Delta}(q^2)}{M_N^{2}}q_\lambda q_\nu \bigg]N(p,s)
\label{matrixelement}
\end{eqnarray}
where $M_N$ is the mass of the nucleon, and $q=p-p'$.

The remaining matrix element, that of the interpolating current $\langle 0|J^\Delta_{\mu}|\Delta(p',s')\rangle$ is defined as
$$\langle 0|J^\Delta_{\mu}|\Delta(p',s')\rangle=\lambda_\Delta\upsilon_\mu(p',s')$$
where $\lambda_\Delta$ is overlap amplitude of $\Delta$ baryon and $\upsilon_\mu(p',s')$ is the $\Delta$  baryon spinor.
Summation over spins of $\Delta$ baryon is defined as:
\begin{equation}\label{spinor}
\sum_{s'}\upsilon_{\mu}(p',s')\overline{\upsilon_{\nu}}(p',s')=-(\not\!p'+M_{\Delta})\bigg[g_{\mu\nu}-\frac{1}{3}\gamma_{\mu}\gamma_{\nu}-\frac{2p'_{\mu}p'_{\nu}}
{3M_{\Delta}^{2}}+\frac{p'_{\mu}\gamma_{\nu}-p'_{\nu}\gamma_{\mu}}{3M_{\Delta}}\bigg]
\end{equation}
Inserting   Eq. (\ref{matrixelement}) into Eq. (\ref{fiz1}) and using Eq. (\ref{spinor}), the correlation function Eq. (\ref{corrf}) can be expressed in terms of the form factors as follows:
  \begin{align}
   \Pi_{\mu\nu}(p,q)&=-i \frac{\lambda_{\Delta}}{M_\Delta^2-p'^{2}} \bigg[C_3^{N \Delta}(q^2)\left\{1-\frac{M_{\Delta}}{M_N}\right\}(q_\mu\gamma_\nu
   - g_{\mu\nu} \qslash)+\left\{C_5^{N \Delta}(q^2)+C_4^{A}(q^2)\frac{p'.q}{M^2_N}\right\} g_{\mu\nu}\qslash\nonumber\\
           &+\frac{C_3^{N \Delta}(q^2)}{M_N} (q_\mu\gamma_\nu-g_{\mu\nu}\qslash) \qslash
		+ \left\{-\frac{2C_3^{N \Delta}(q^2)}{M_N}-\frac{C_4^{N \Delta}(q^2)}{M_N}(1+\frac{M_{\Delta}}{M_N})\right\}(q_\mu p'_\nu-g_{\mu\nu}p'.q)\nonumber\\
         &+\left\{C_5^{N \Delta}(q^2)(M_{\Delta}+M_N)\right\}g_{\mu\nu}
 		 +\left\{C_6^{N \Delta}(q^2)(\frac{M_{\Delta}+M_N}{M_N^2})\right\}q_\mu q_\nu
 		 -\left\{ \frac{C_4^{N \Delta}(q^2)}{M_N^2}\right\}q_\mu p'_\nu \qslash \nonumber\\
 		 &+\left\{ \frac{C_6^{N \Delta}(q^2)}{M_N^2}\right\}q_\mu q_\nu \qslash \bigg]
\label{hadronic}
  \end{align}

In this expression, only the contribution of the spin-$3/2$ $\Delta$ baryon is shown.
In principle, the correlation function can also receive contributions from spin-$1/2$ particles.
The overlap of the spin-$1/2$ particles with the $J_\mu^\Delta$ current can be written as
\begin{eqnarray}
\langle 1/2(p') \vert J^\Delta_\mu \vert 0 \rangle = \left(A p'_\mu + B \gamma_\mu \right) u(p')
\end{eqnarray}
where $u(p')$ is the spinor describing the spin-$1/2$ particle. Hence, if in the correlation function,
the $\gamma_\mu$ matrix is carried to the left, only the terms that are proportional to $p'_\mu$ or contain a $\gamma_\mu$ at the far
left receive contributions from the spin-$1/2$ particles. Following this observation,
we will order the gamma matrices as $\gamma_\mu \gamma_\nu\not\!q \not\!p'$. The $\not\!p'$ matrix can be eliminated using the equation of motion.
After these steps, any structure that is not proportional to $p'_\mu$,
or that does not contain a $\gamma_\mu$ receives contributions from the spin-$3/2$ particles only \cite{Belyaev:1993ss, Belyaev:1982cd}.

To obtain the expression of the correlation function in terms of the QCD parameters and the DAs,
an explicit form for the  interpolating current of the $\Delta$ baryon needs  to be chosen. In this work, the interpolating current is chosen as follows:
 	\begin{eqnarray}\label{intf}
		J_{\mu}^{\Delta}(x)=&\frac{1}{\sqrt{3}}\epsilon^{abc}[2(u^{aT}(x) C\gamma_\mu d^b(x)) u^c(x)+ (u^{aT}(x)C\gamma_\mu u^b(x))d^c(x)]
	\end{eqnarray}
Here  $a$, $b$, $c$ are  color indices and $C$ denotes charge conjugation.
After  contracting one pair of the light quark operators, the correlation function becomes:
 \begin{eqnarray} \label{eq:Qcds}
\left( \Pi_{\mu\nu} \right)_{\lambda\eta}&=&\frac{i}{8\sqrt3}\int d^4 x e^{iqx}~(C\gamma_\mu)_{\alpha\beta}(\gamma_\nu \gamma_5)_{\rho\sigma}\left\{4\epsilon^{abc}\langle \nonumber
0|{q_1}_{\sigma}^a(0) {q_2}_{\theta}^b(x) {q_3}_{\phi}^c(0)|N(p,s) \rangle \right. \\ \nonumber
& &\left.\ \bigg[2\delta_\alpha^{\eta}\delta_\sigma^{\theta} \delta_\beta^{\phi}S(-x)_{\lambda\rho}+
2\delta_\lambda^{\eta} \delta_\sigma^{\theta} \delta_\beta^{\phi}S(-x)_{\alpha\rho}\right.\nonumber
\left.+ \delta_\alpha^{\eta} \delta_\sigma^{\theta} \delta_\lambda^{\phi} S(-x)_{\beta\rho}+\delta_\beta^{\eta} \delta_\sigma^{\theta} \delta_\phi^{\lambda} S(-x)_{\alpha\rho}\bigg]\right.\nonumber \\
& &\left.
- 4\epsilon^{abc}\langle 0|{q_1}_{\sigma}^a(0) {q_2}_{\theta}^b(0) {q_3}_{\phi}^c(x)|N(p,s) \rangle \right.
\left.\bigg[ 2\delta_\alpha^{\eta}\delta_\lambda^{\theta} \delta_\sigma^{\phi} S(-x)_{\beta\rho}
+\delta_\alpha^{\eta}\delta_\beta^{\theta} \delta_\sigma^{\phi} S(-x)_{\lambda\rho}\bigg] \right\}
	\label{eq9}
\end{eqnarray}
where $\lambda$ and $\eta$ are spinor indices, and $S(x)$ represents the light-quark propagator and is given by: 
\begin{eqnarray}\label{pro}
	S_q(x)&=&\frac{i\xslash}{2\pi^2x^4}-\frac{\langle q\bar{q}\rangle}{12}\left(1+\frac{m_0^2 x^2}{16}\right)-ig_s\int^1_0 d\upsilon\left[\frac{\xslash}{16\pi^2x^4} G_{\mu\nu}\sigma^{\mu\nu}-\upsilon x^\mu G_{\mu\nu}\gamma^\nu\ \frac{i}{4\pi^2x^2}\right].
\end{eqnarray}
In this expression, $\langle q \bar q \rangle$ is the quark condensate, $m_0$ is defined in terms of the mixed quark gluon condensate as $\langle \bar q g_s G^{\mu \nu} \sigma_{\mu \nu} q\rangle \equiv  m_0^2 \langle \bar q q \rangle$ and $g_s$ is the strong coupling constant.
The terms proportional to $G_{\mu\nu}$ arise from the interaction of the propagating quark with the external gluonic field and  lead to  contributions from the four-particle
nucleon distribution amplitudes.
Such corrections from higher Fock space components of the nucleon wave function are not expected to play an important role~\cite{Diehl:1998kh},
and hence  we would not take them into account. Additionally, the terms proportional to quark condensates 
are removed by Borel transformations.
Hence the first term, which contains the hard light-quark propagator, will be considered for our discussion.	
To proceed with the calculation of the correlation function, the matrix element of the local three-quark operator
$4\epsilon^{abc}\langle 0|q_{1\alpha}^a(a_1 x) q_{2\beta}^b(a_2 x) q_{3\gamma}^c(a_3 x)|N(p,s)\rangle$
is needed.
The light-cone distribution amplitudes of the nucleon, which we use in our work to extract the axial form factors,
 are presented in
Ref.~\cite{Braun:2006hz} up to twist six on the basis of QCD conformal partial wave expansion. The expansion is in terms of increasing twist, where the twist of a DA is defined as the difference between the dimension and spin of the operators contributing to that DAs. 
We refer the reader to Refs.~\cite{Braun:2006hz} for a detailed analysis of the  distribution amplitudes of the nucleon.
Using the most general decomposition of the matrix element~(see Eq.~(2.3) in Ref.~\cite{Braun:2000kw}) and taking the Fourier transformations appearing in Eq. (\ref{eq9}), the QCD representation of the correlation function can be obtained.

Note that the hadronic representation, Eq. (\ref{hadronic}), and the QCD representation are obtained in different kinematical regions. The two expression can be related to each other by using the spectral representation of the correlation functions. Quite generally, the coefficients of various structures in the correlation function can be written as:
\begin{eqnarray}
\Pi(p^2,{p'}^2; Q^2) = \int_0^\infty ds_1 ds_2 \frac{\rho(s_1,s_2;Q^2)}{(s_1-p^2)(s_2-{p'}^2)} + \mbox{polynomials in $p^2$ or ${p'}^2$}
\end{eqnarray}
where $\rho$ is called the spectral density. The spectral density can be calculated both using the hadronic representation of the correlation function, $\rho^h$, or using the QCD representation, $\rho^{QCD}$. Once $\rho$ is obtained, the spectral representation allows one to evaluate the correlation function in all kinematical regions for $p^2$ and ${p'}^2$.

The LCSR are obtained by matching the expression of the correlation function in terms of QCD parameters to its expression in terms of the hadronic properties,
using their spectral representation.
In order to do this, we choose the structures proportional to $ q_\mu\gamma_\nu\qslash$,
$q_\mu{p'}_\nu\qslash$, $(g_{\mu\nu}\qslash - q_\mu\gamma_\nu\qslash)$ and $q_\mu q_\nu\qslash$ for the form factors
 $C_3^{N\Delta}$, $C_4^{N\Delta}$, $C_5^{N\Delta}$ and $C_6^{N\Delta}$, respectively.
For the $N\rightarrow \Delta$ transition form factors, we obtain:

\begin{align}
C_3^{N \Delta}(Q^2) \frac{\lambda_{\Delta}}{M_\Delta^2-p'^{2}}&=
					- \frac{M^{3}_{N}}{ \sqrt{3}}\int_{0}^{1} {d\alpha} \frac{(1-\alpha)}{(q-p\alpha)^{4}} [F_9(\alpha)-F_{10}(\alpha)]
					 + \frac{M_N}{\sqrt{3}}\int_0^{1}{dx_i}\frac{1}{(q-px_i)^2} [2F_{11}(x_i)-F_{12}(x_i)] \nonumber \\
					 &-\frac{M^{3}_{N}}{ \sqrt{3}}\int_0^{1}{dx_i}\frac{1}{(q-px_i)^4}[2F_{13}(x_i)-F_{14}(x_i)]
					 -\frac{M_N^{^3}}{\sqrt{3}}\int_0^{1}{d\beta}\frac{1}{(q-p\beta)^{4}} [2F_{15}(\beta)-F_{16}(\alpha)]
 \label{eq:ffc3}
\\
C_4^{N \Delta}(Q^2) \frac{\lambda_{\Delta}}{M_\Delta^2-p'^{2}}&=
					- \frac{M^{5}_{N}}{ \sqrt{3}}\int_{0}^{1}{d\beta} \frac{(1-\beta)^{2}} {(q-p\beta)^{6}}	[4F_{5}({\beta})- 2F_{7}({\beta})]
					+\frac{M_{N}^{3}}{\sqrt{3}}\int_{o}^{1}{d\alpha} \frac{(1-\alpha)}{(q-p\alpha)^{4}} [2F_{6}(\alpha)+F_{8}(\alpha)]
\label{eq:ffc4}
\end{align}
\begin{align}
C_5^{N \Delta}(Q^2)  \frac{\lambda_{\Delta}}{M_\Delta^2-p'^{2}} &=
				 \frac{M^{3}_{N}}{ \sqrt{3}}\int_{0}^{1}{d\beta} \frac{1} {(q-p\beta)^{4}}[F_{17}(\beta)-F_{18}(\beta)]
				 - \frac{M^{3}_{N}}{ \sqrt{3}}\int_0^{1}dx_i \frac{1} {(q-p{x_i})^{4}}[F_{19}(x_i)+F_{20}(x_i)]\nonumber\\
				 &+\frac{M_{N}}{ \sqrt{3}}\int_{0}^{1}{dx_i} \frac{1} {(q-p{x_i})^{2}}[F_{21}(x_i)-F_{22}(x_i)]
				 - \frac{M^{3}_{N}}{ \sqrt{3}}\int_{0}^{1}{d\beta} \frac{2} {(q-p\beta)^{4}}[F_{23}(\beta)-F_{24}(\beta)]\nonumber\\
			         &+\frac{M_{N}}{ \sqrt{3}}\int_{0}^{1}{dx_i} \frac{1-{x_i}} {(q-p{x_i})^{2}}[F_{25}(x_i)-F_{26}(x_i)]
			         - \frac{M^{3}_{N}}{ \sqrt{3}}\int_{0}^{1}{dx_i} \frac{1} {(q-p{x_i})^{4}}[F_{27}(x_i)-F_{28}(x_i)]\nonumber\\
				 &+ \frac{M^{3}_{N}}{ \sqrt{3}}\int_{0}^{1}{d\beta} \frac{1-\beta} {(q-p\beta)^{4}}[F_{29}(\beta)-F_{30}(\beta)]
				 + \frac{M^{3}_{N}}{ \sqrt{3}}\int_{0}^{1}{d\alpha} \frac{1-\alpha} {(q-p\alpha)^{4}}[F_{31}(\alpha)-F_{32}(\alpha)]\nonumber\\
				 &-\frac{M_{N}}{ \sqrt{3}}\int_{0}^{1}{dx_i} \frac{1} {(q-p{x_i})^{2}}[2F_{33}(x_i)-F_{34}(x_i)]
				 + \frac{M^{3}_{N}}{ \sqrt{3}}\int_0^{1}dx_i \frac{1} {(q-p{x_i})^{4}}[2F_{35}(x_i)+F_{36}(x_i)]\nonumber\\
				 &+ \frac{M^{3}_{N}}{ \sqrt{3}}\int_{0}^{1}{d\beta} \frac{1} {(q-p\beta)^{2}}[2F_{37}(\beta)-F_{38}(\beta)]
				 + \frac{M^{3}_{N}}{ \sqrt{3}}\int_{0}^{1}{d\beta} \frac{1-\beta} {(q-p\beta)^{4}}[2F_{39}(\beta)+F_{40}(\beta)]\nonumber\\
				 &- \frac{M_{N}}{ \sqrt{3}}\int_{0}^{1}{d\alpha} \frac{1} {(q-p\alpha)^{2}}[F_{41}(\alpha)+F_{42}(\alpha)]
\label{eq:ffc5}
\end{align}
\begin{align}
C_6^{N \Delta}(Q^2) \frac{\lambda_{\Delta}}{M_\Delta^2-p'^{2}}&=
                   		\frac{M^{5}_{N}}{ \sqrt{3}}\int_0^{1}d\beta \frac{(1-\beta)^{2}} {(q-p\beta)^{6}} [4F_{1}(\beta)+2F_{3}(\beta)]
                   		  - \frac{M_{N}^{3}}{\sqrt{3}}\int_{o}^{1}{d\alpha} \frac{(1-\alpha)}  {(q-p\alpha)^{4}}[F_{2}(\alpha)-F_{4}(\alpha)]                  	
\label{eq:ffc6}
\end{align}
The explicit forms of the functions that appear in the above sum rules in terms of the DAs of the nucleon are given in appendix A.
In these expression, the left hand side is actually a sum over the contributions of all spin-$3/2$ baryons. The sum rules is obtained by carrying out Borel transformation to eliminate any polynomials that arise during the matching. Furthermore, Borel transformation also suppresses the contributions of higher states and continuum.
After Borel transformations, Eqs. (\ref{eq:ffc3}-\ref{eq:ffc6}) takes the form
\begin{equation}
C_i^{N\Delta}(Q^2) e^{-\frac{m_\Delta^2}{M^2}} = \int_0^1 dx \rho^i(x) e^{-\frac{s(x)}{M^2}}~~~~ i=3,~4,~5, \mbox{ or } 6
\label{eq:sr}
\end{equation}
where
$$s(x)=(1-x)M_N^2+\frac{1-x}{x}Q^2,$$
with $Q^2=-q^2$ and $M$ is the Borel parameter.

\section{Traditional Analysis vs. Monte Carlo Analysis}
In the traditional analysis of sum rules, the spectral density of the higher states and the continuum are parameterised using quark hadron duality.
In this approach, it is assumed that
$\rho^h(s) = \rho^{QCD}(s)$ when $s > s_0$, i.e. the contribution of the higher states and continuum to the spectral density is approximated by the spectral density expressed in terms of the QCD parameters.
Both  the Borel transformation and the subtraction of the higher states and the continuum are carried out using the following
substitution rules (see e.g. \cite{Braun:2006hz}):
	\begin{align}\label{subtract}
		&\int dx \frac{\rho(x)}{(q-xp)^2}\rightarrow -\int_{x_0}^1\frac{dx}{x}\rho(x) e^{-s(x)/M^2}, \nonumber\\
\nonumber\\
		&\int dx \frac{\rho(x)}{(q-xp)^4}\rightarrow \frac{1}{M^2} \int_{x_0}^1\frac{dx}{x^2}\rho(x) e^{-s(x)/M^2}
		+\frac{\rho(x_0)}{Q^2+x_0^2 M_N^2} e^{-s_0/M^2},\nonumber\\
\nonumber\\
        &\int dx \frac{\rho(x)}{(q-xp)^6}\rightarrow -\frac{1}{2M^4}\int_{x_0}^1\frac{dx}{x^3}\rho(x) e^{-s(x)/M^2}
        -\frac{1}{2M^2}\frac{\rho(x_0)}{x_0(Q^2+x_0^2M_N^2)}e^{-s_0/M^2}\nonumber\\
         \nonumber\\
	\hspace{2cm}	&\qquad\qquad\qquad\qquad+\frac{1}{2}\frac{x_0^2}{Q^2+x_0^2M_N^2}\bigg[\frac{d}{dx_0}\frac{\rho(x_0)}{x_0(Q^2+x_0^2M_N^2)}\bigg]e^{-s_0/M^2},
\end{align}
where
 $x_0$ is the solution of the quadratic equation for $s=s_0$:
\[x_0=\left[\sqrt{(Q^2+s_0-M_N^2)^2+4M_N^2 Q^2}-(Q^2+s_0-M_N^2)\right]/(2M_N^2),\] where $s_0$ is the continuum threshold. $s_0 \rightarrow \infty$ ($x_0=0$) limit corresponds to Borel transformation without subtraction.

An alternative approach to analyse the sum rules using Monte Carlo methods is presented in \cite{Leinweber:1995fn}. The method  has the advantage that it allows for a more reliable estimate of the error bars.
 In this method, one chooses random values for the uncertain parameters appearing
in the sum rules within their error bars. For each set of values of the parameters, one obtains a numerical prediction for quantity under study.
The obtained results are than analysed statistically to obtain the central value and the error bars ( see \cite{Leinweber:1995fn}
for details). In \cite{Leinweber:1995fn}, the method has been used for mass sum rules, and in \cite{Lee:1996dc, Wang:2009ru}, it has been applied to calculating coupling constants. Here we generalize that method to analyze the form factors. The steps of the analysis can be summarised as follows:
\begin{enumerate}
\item Let $p_i$ denote one of the input parameters whose central value is $\bar p_i$ and whose error is $\sigma_i$. We will assume that $p_i$ is a normally distributed random variable with variance $\sigma_i$. Let $\{p_k\}$ denote a possible set of input parameters. Choose $N$ such sets. In this work, $N$ is chosen to be $N=1000$.
\item Let $\Pi_{\{p_k\}}(Q^2,M^2)$  be the correlation function calculated using the set $\{p_k\}$ of input parameters for the fixed values of $M^2$ and $Q^2$. Denote the average and the standard deviation of these values by $\bar \Pi(Q^2,M^2)$ and $\sigma_\Pi(Q^2,M^2)$.
\item For each set $\{p_k\}$ of parameters,  define $\chi^2_{\{p_k\}}$ distribution for a fixed $Q^2$ as
\begin{equation}
\chi^2_{\{p_k\}}(Q^2) = \sum_{M^2} \frac{\left(\Pi_{\{p_k\}}(Q^2,M^2) - \Pi^{model}(Q^2,M^2;C(Q^2),a_n) \right)^2}{\sigma_\Pi(Q^2,M^2)^2}
\end{equation}
where the model for the correlation function depends on the form factor $C(Q^2)$ and possibly other parameters $a_n$.
\item Minimizing $\chi^2_{\{p_k\}}(Q^2)$ with respect to the form factor $C(Q^2)$ and other parameter of the model for the correlation function yields the form factor for the given set of parameter $C_{\{p_k\}}(Q^2)$.
\item Repeating this procedure for all the parameter sets, a distribution for the value of the form factor at the chosen value of $Q^2$ is obtained. The prediction for the value of the form factor at the chosen value of $Q^2$ can be taken as the average
of this distribution and the uncertainty as the standard deviation of this distribution.
\item This procedure is repeated for various values of $Q^2$ to obtain the value and the uncertainty of the value
of the form factor can be obtained.
\end{enumerate}

In this work, the correlation has been modelled as
\begin{equation}
\Pi = a_0 C^{N\Delta}(Q^2) e^{-\frac{M_\Delta^2}{M^2}} + a_1 e^{-\frac{m_1^2}{M^2}} + a_2 e^{-\frac{m_2^2}{M^2}}
\end{equation}
where we imposed the constraint $m_2>m_1>M_\Delta$. Both a triple exponential (TE) and a double exponential (DE) fit (by setting $a_2=0$) has been performed.

\section{Results and Conclusion}
To obtain a numerical prediction for the form factors, first one needs the expressions for  the distribution amplitudes of the nucleon \cite{Braun:2006hz}, 
the distribution amplitudes can be expressed in terms of several parameter. Not all of the parameters are independent. The independent parameters and their numerical values are presented in
Table \ref{parameter_table} \cite{Braun:2006hz}.
Another non-perturbative parameter that is required is the residue of the $\Delta$ baryon $\lambda_\Delta$.
This residue is obtained from the mass sum rules for the $\Delta$ baryons to be $\lambda_\Delta= 0.038 ~GeV^3$
\cite{Aliev:2004ju, Lee:1997jk, Hwang:1994vp}.
\begin{table}[!h]
	\addtolength{\tabcolsep}{10pt}
\begin{tabular}{ccccccc}
		\hline\hline
		& $f_N$~(GeV$^2$) & $\lambda_1$~(GeV$^2$)& $\lambda_2$~(GeV$^2$)& \\[0.5ex]
		\hline
		 & 0.005 $\pm$ 0.0005 &-0.027$\pm$ 0.009 & 0.054$\pm$ 0.019\\[0.8ex]
			\hline	\hline
		 $V_1^d$ & $A_1^u$ & $f_1^d$ & $f_2^d$ & $f_1^u$  \\[0.5ex]
		\hline
		 0.23 $\pm$ 0.03 & 0.38$\pm$ 0.15& 0.40 $\pm$ 0.05 & 0.22$\pm$ 0.05 & 0.07$\pm$ 0.05 &  \\[0.5ex]
		\hline\hline
	\end{tabular}
	\caption{The values of the parameters entering the DAs of $N$.
	The upper panel shows the dimensionful parameters N.
	In the lower panel we list the values of the five parameters that determine the shape of the DAs.}
	\label{parameter_table}
\end{table}

\begin{figure}[htp]
\centering
\subfloat[]{\label{fig:c3MDsq.pdf}
\includegraphics[width=0.4\textwidth]{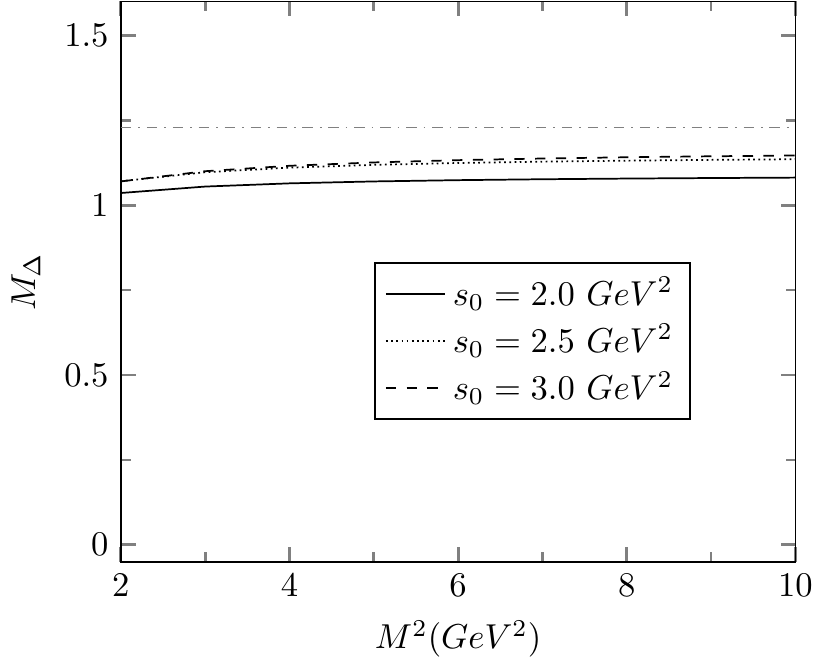}}
  \subfloat[]{\label{fig:c4MDsq.pdf}
  \includegraphics[width=0.4\textwidth]{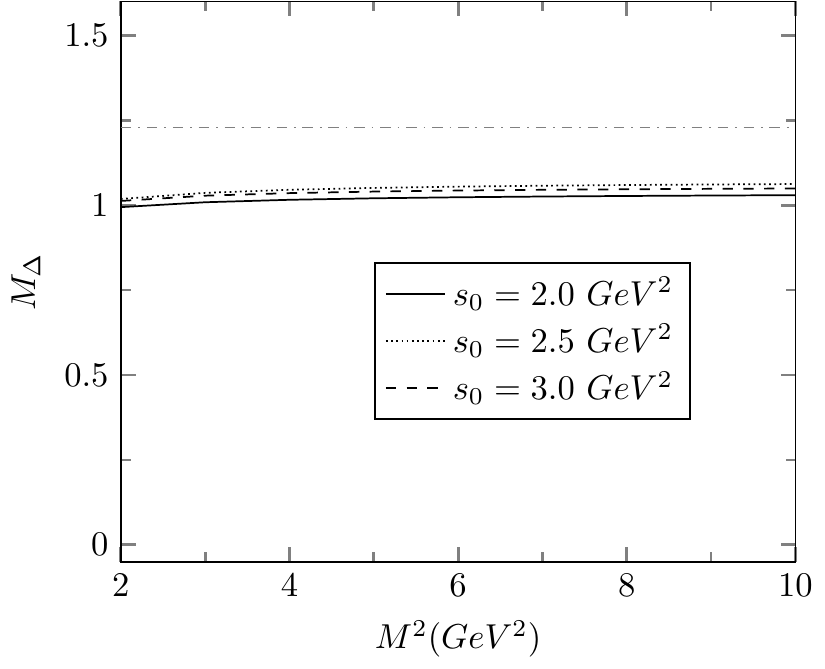}}
  \\
   \subfloat[]{\label{fig:c5MDsq.pdf}
   \includegraphics[width=0.4\textwidth]{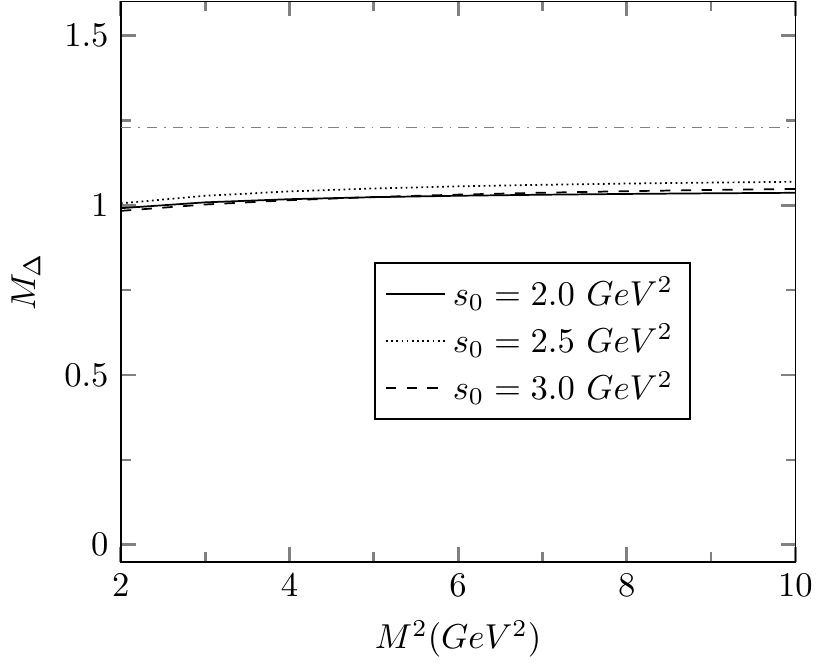}}
    \subfloat[]{\label{fig:c6MDsq.pdf}
    \includegraphics[width=0.4\textwidth]{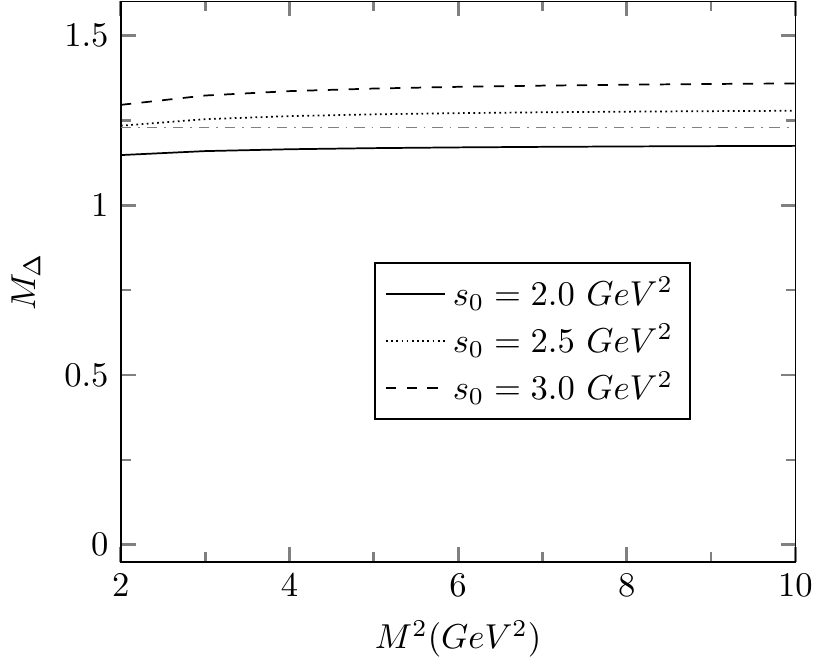}}
\caption{The $M_\Delta$ predictions obtained from the form factor sum rules for $s_0=2.0~GeV^2$, $s_0=2.5~GeV^2$ and $s_0=3.0~GeV^2$ at $Q^2=2.0~GeV^2$, 
(a) for $C_3^{N \Delta}$ form factors,
(b) for $C_4^{N \Delta}$ form factors,
(c) for $C_5^{N \Delta}$ form factors,
(d) for $C_6^{N \Delta}$ form factors.}
\label{fig:MD}
\end{figure}

 In the conventional analysis, the obtained predictions  for the form factors
depend on two auxiliary parameters: the Borel parameter $M^2$, and the continuum threshold $s_0$.
The continuum threshold signals the energy the scale at which, the excited states and continuum start to contribute to the correlation function. There are various proposal on how to determine this parameter.  One approach is to vary this parameter in a reasonable range, until a Borel window appears in which the predictions are independent of the Borel parameter \cite{Ligeti:1993qd}. Another recent proposal is to choose $s_0$ as a function of the Borel parameter and to determine the functional dependence by requiring the independence of the mass prediction on the Borel parameter \cite{Lucha:2009uy}. The mass of the $\Delta$ baryon can be obtained from Eq. (\ref{eq:sr}) using the relation:
\begin{align}
M_\Delta^2 = M^4 \frac{\partial}{\partial M^2} \ln \left(C_i^{N\Delta}(Q^2) e^{-\frac{M_\Delta^2}{M^2}} \right)
\end{align}
But the generally accepted rule of thumb to determine this parameter is to assume that $s_0\simeq(M_\Delta+0.3~GeV)^2$ and check the dependence of the results on
slight variations of this parameter. To estimate the reliability of this determination of $s_0$, in Fig. (\ref{fig:MD}), we present the $M_\Delta$ predictions obtained from the form factor sum rules for $s_0=2.0~GeV^2$, $s_0=2.5~GeV^2$ and $s_0=3.0~GeV^2$ at $Q^2=2.0~GeV^2$. As can be seen from the figures, the predictions for the mass of $M_\Delta$ are within about $20\%$ for all of the sum rules. Also, as can also be seen from the figures, within the considered range of $M^2$, the mass predictions are almost independent of the Borel parameter. Hence, a possible $M^2$ dependence of $s_0$ is not necessary within this range.

\begin{figure}[htp]
\centering
 \subfloat[]{\label{fig:NC3Msq.eps}\includegraphics[width=0.3\textwidth]{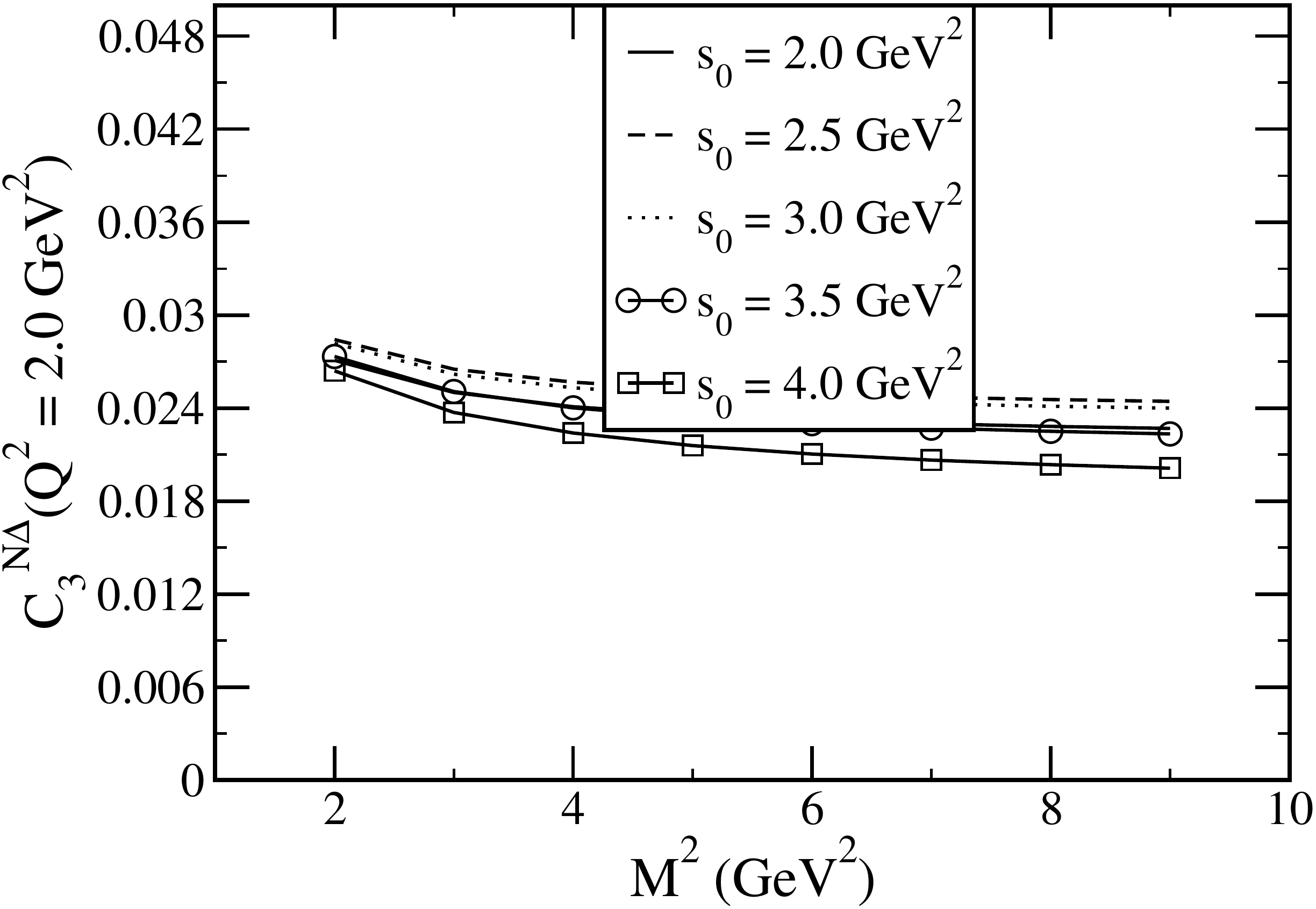}}
 \subfloat[]{\label{fig:NC3Msq1.eps}\includegraphics[width=0.3\textwidth]{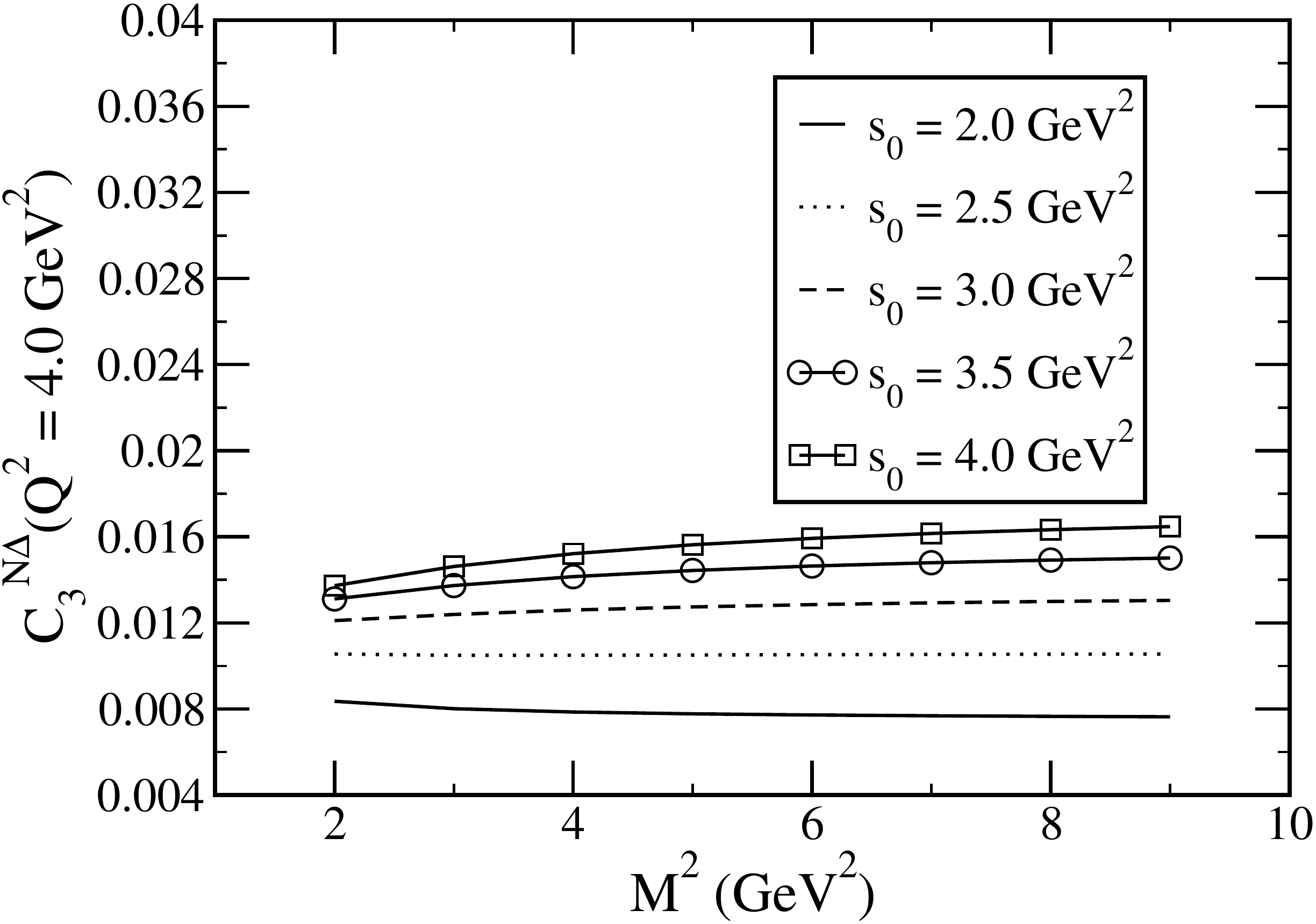}}\\
 \subfloat[]{\label{fig:NC4Msq.eps}\includegraphics[width=0.3\textwidth]{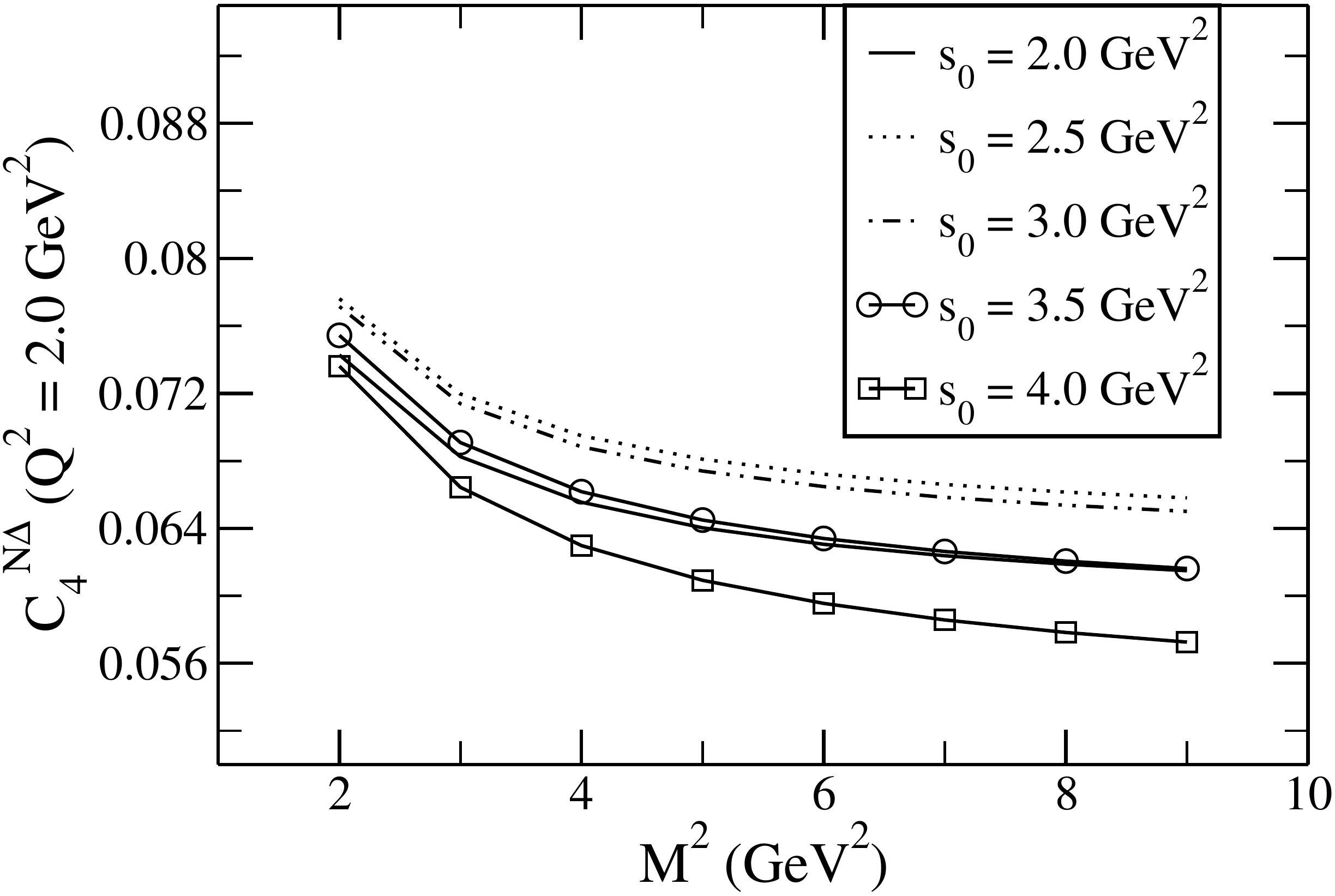}}
 \subfloat[]{\label{fig:NC4Msq1.eps}\includegraphics[width=0.3\textwidth]{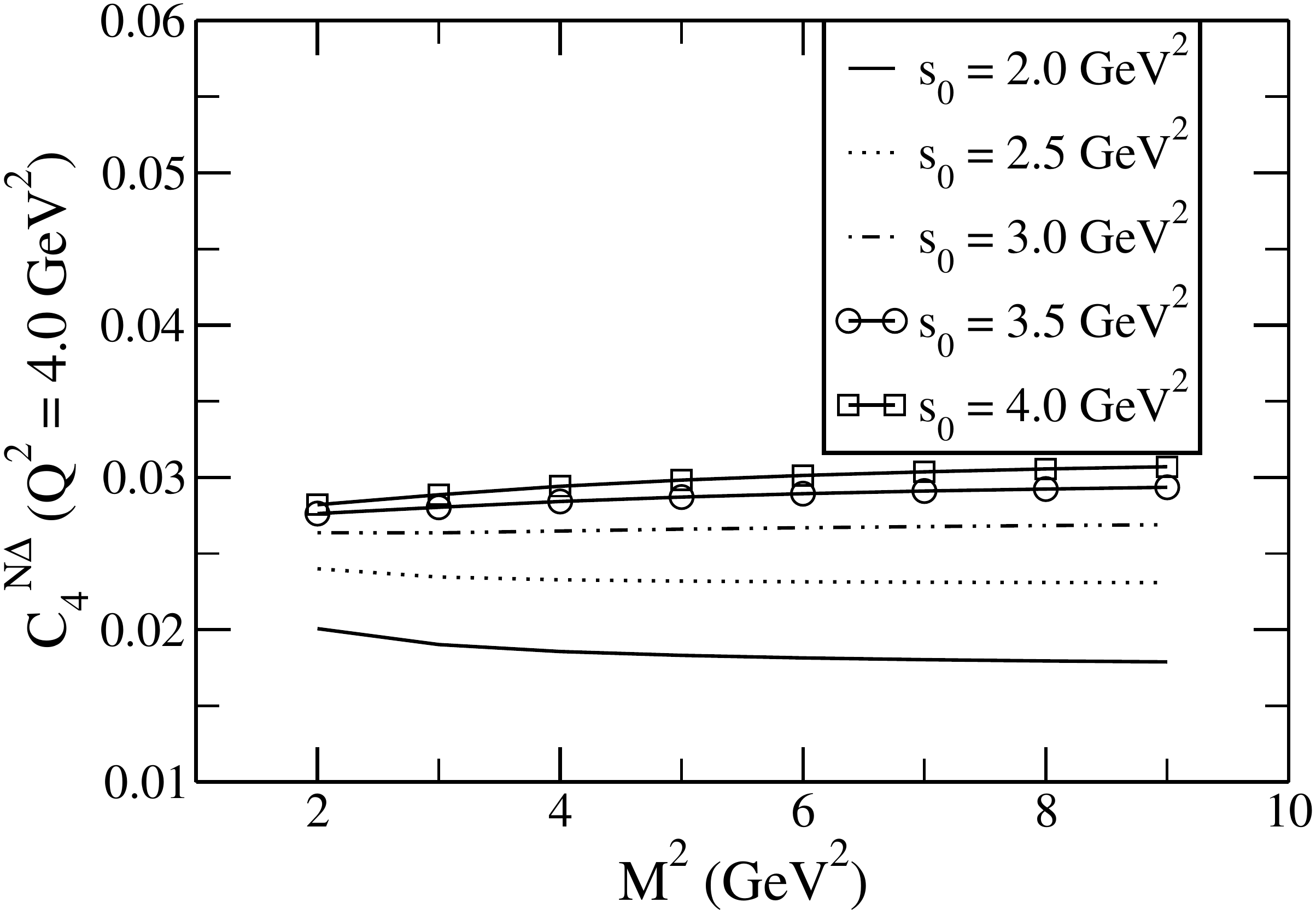}}
 \\
  \subfloat[]{\label{fig:NC5Msq.eps}\includegraphics[width=0.3\textwidth]{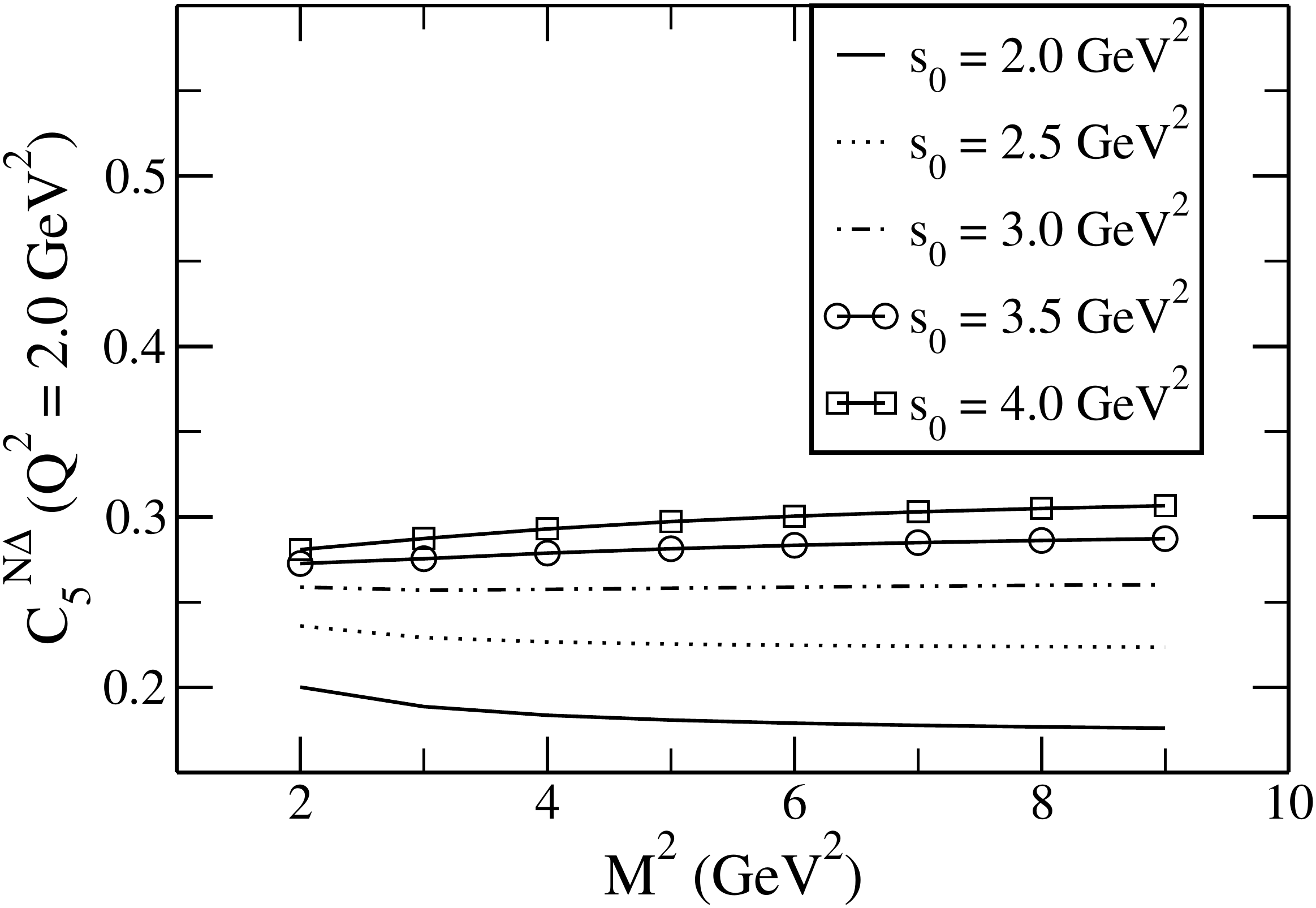}}
  \subfloat[]{\label{fig:NC5Msq1.eps}\includegraphics[width=0.3\textwidth]{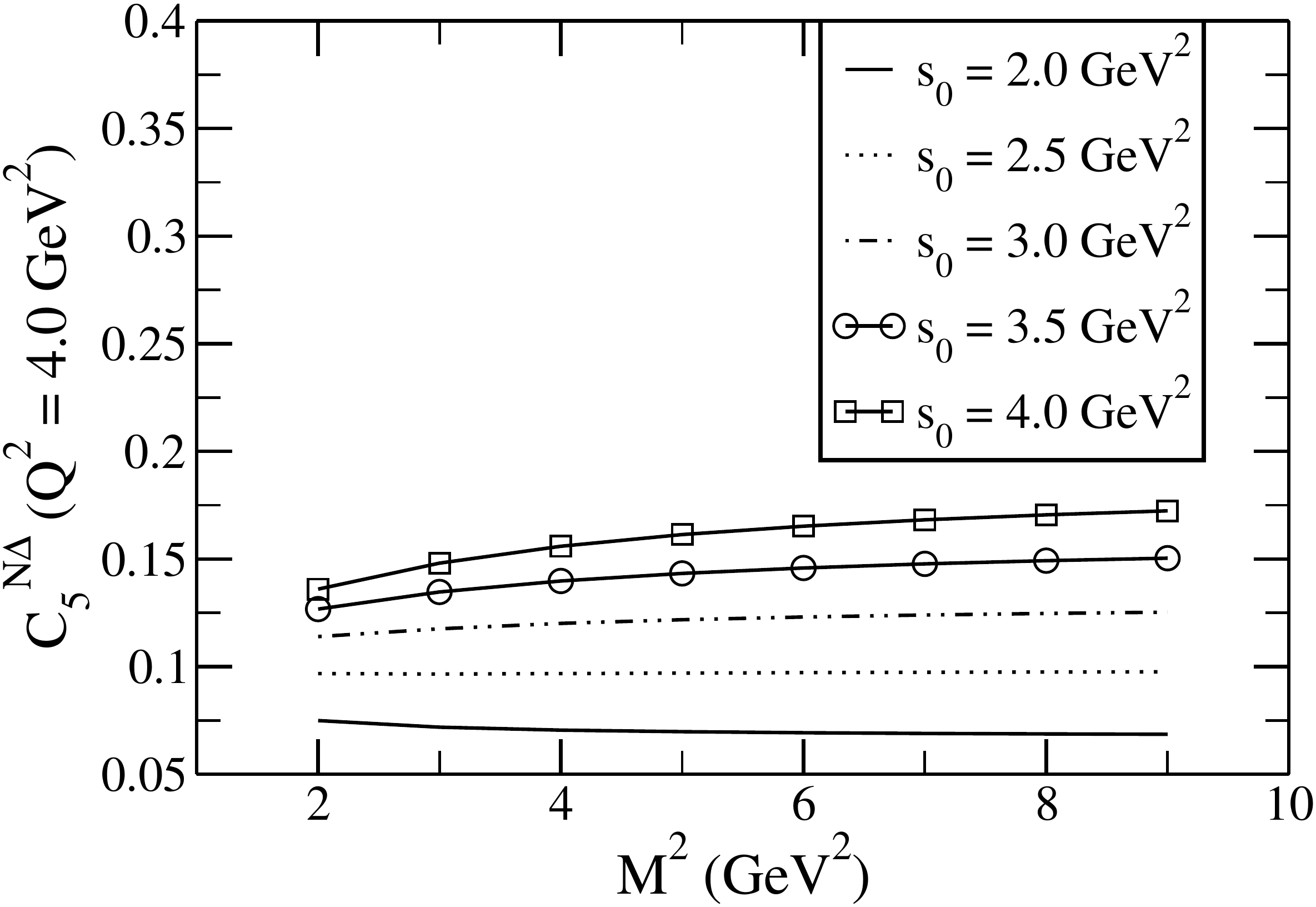}}\\
    \subfloat[]{\label{fig:NC6Msq.eps}\includegraphics[width=0.3\textwidth]{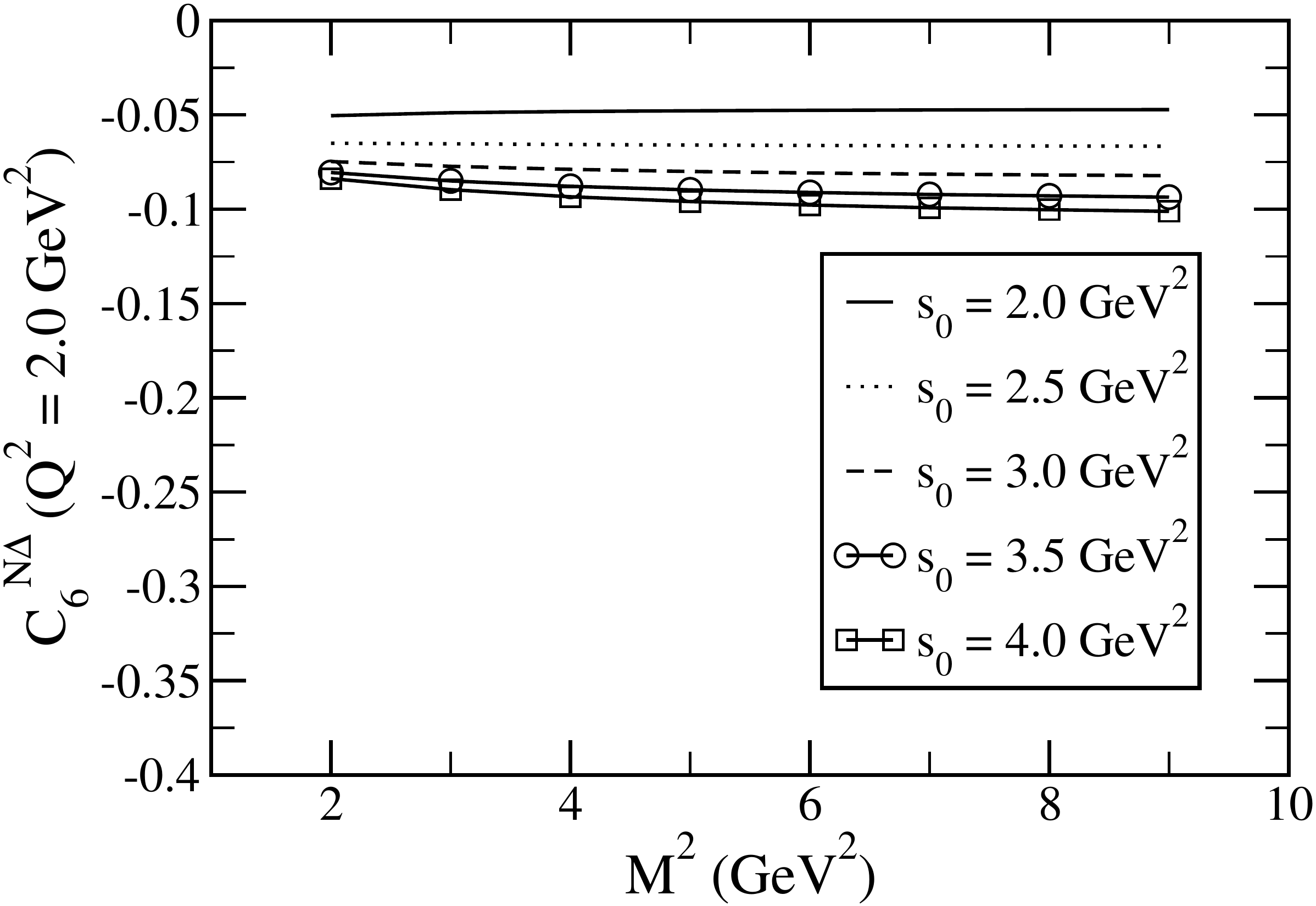}}
    \subfloat[]{\label{fig:NC6Msq1.eps}\includegraphics[width=0.3\textwidth]{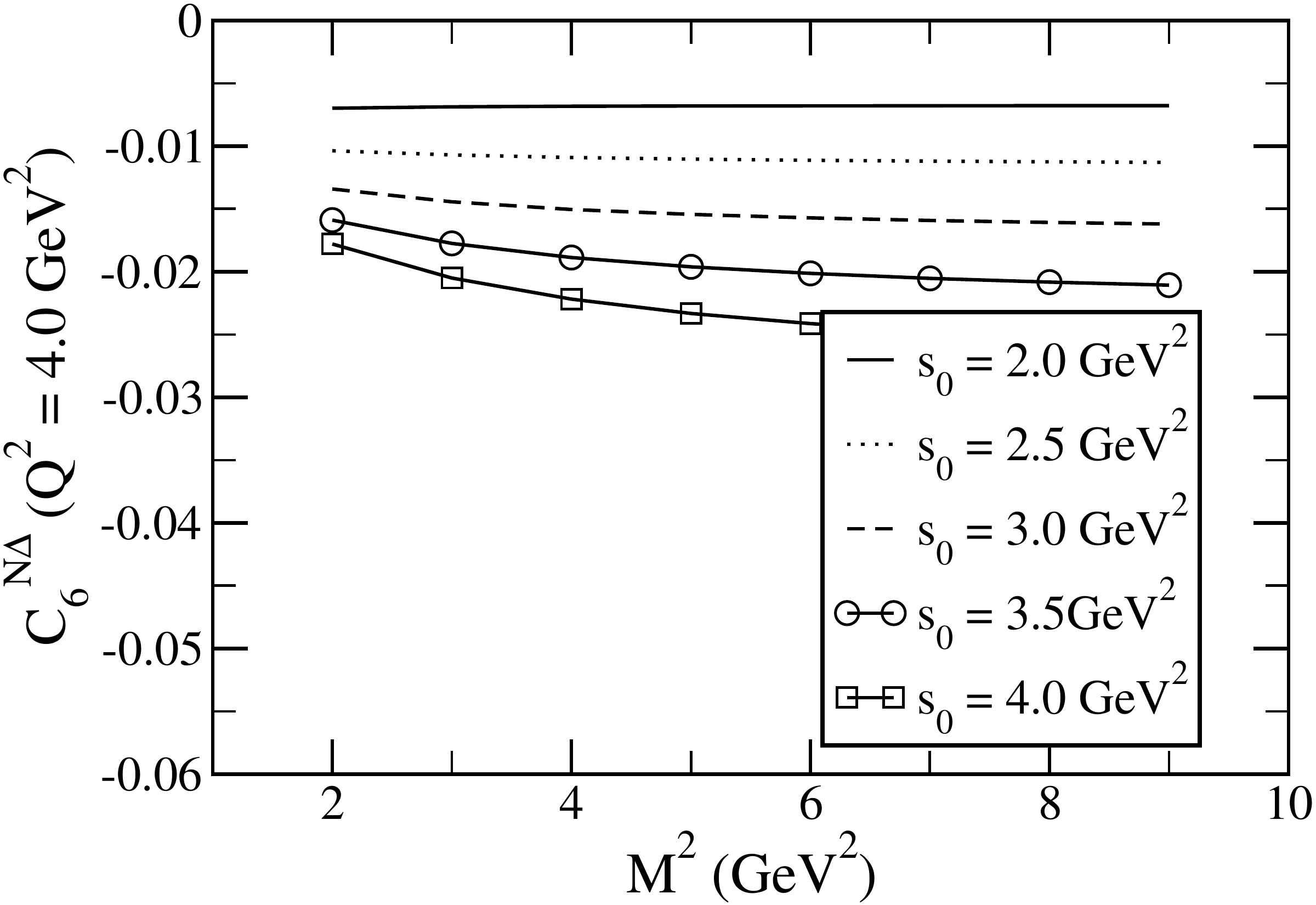}}
\caption{The dependence of the form factors; on the Borel parameter squared $M^{2}$
  for the values of the continuum threshold
$s_0 = 2.0 ~GeV^2$, $s_0 = 2.5 ~GeV^2$, $s_0 = 3.0 ~GeV^2$, $s_0 = 3.5~GeV^2$,  $s_0 = 4.0~GeV^2$ and $Q^2 = 2.0~ GeV^2$,  $4.0~ GeV^2$
(a) and (b) for $C_3^{N \Delta}$ form factor,
(c) and (d) for $C_4^{N \Delta}$ form factor,
(e) and (f) for $C_5^{N \Delta}$ form factor and
(g) and (h) for $C_6^{N \Delta}$ form factor.(Note the different scales used for the vertical axis.)}
\label{msqdependence}
\end{figure}

The Borel parameter is an unphysical parameter, and the predictions on the form factors should be independent of
the value of this parameter. Due to the approximations made to obtain the sum rules, a residual dependence on $M^2$ remains
and for this reason, a region in which the predictions are  practically independent on the value of the Borel parameter needs to be chosen.
In Fig. (\ref{msqdependence}), we plot the dependencies of
the form factors on $M^{2}$ for  two fixed values of $Q^2$ and for various values of $s_0$ in the range $2.0 ~ GeV^2 \leq s_0 \leq 4~GeV^2$.
As can be seen from these figures, for $M^2\ge 3.0~GeV^2$, there is negligible dependence of the Borel parameter.
Hence, in the traditional analysis, this range of $M^2$ is used.

\begin{figure}[htp]
\centering
\subfloat[]{\label{fig:Nc3con.eps}\includegraphics[width=0.4\textwidth]{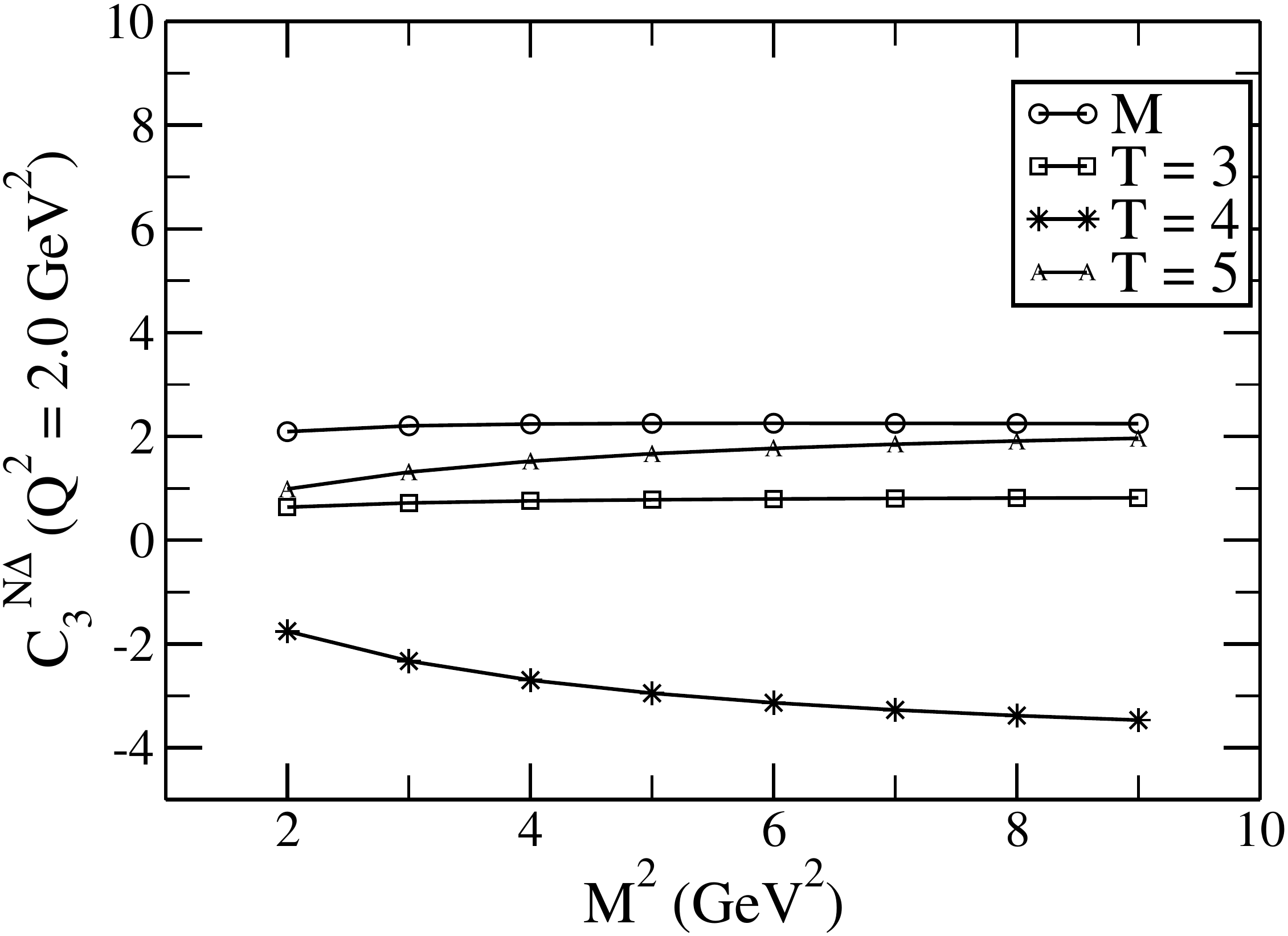}}
  \subfloat[]{\label{fig:Nc4con.eps}\includegraphics[width=0.4\textwidth]{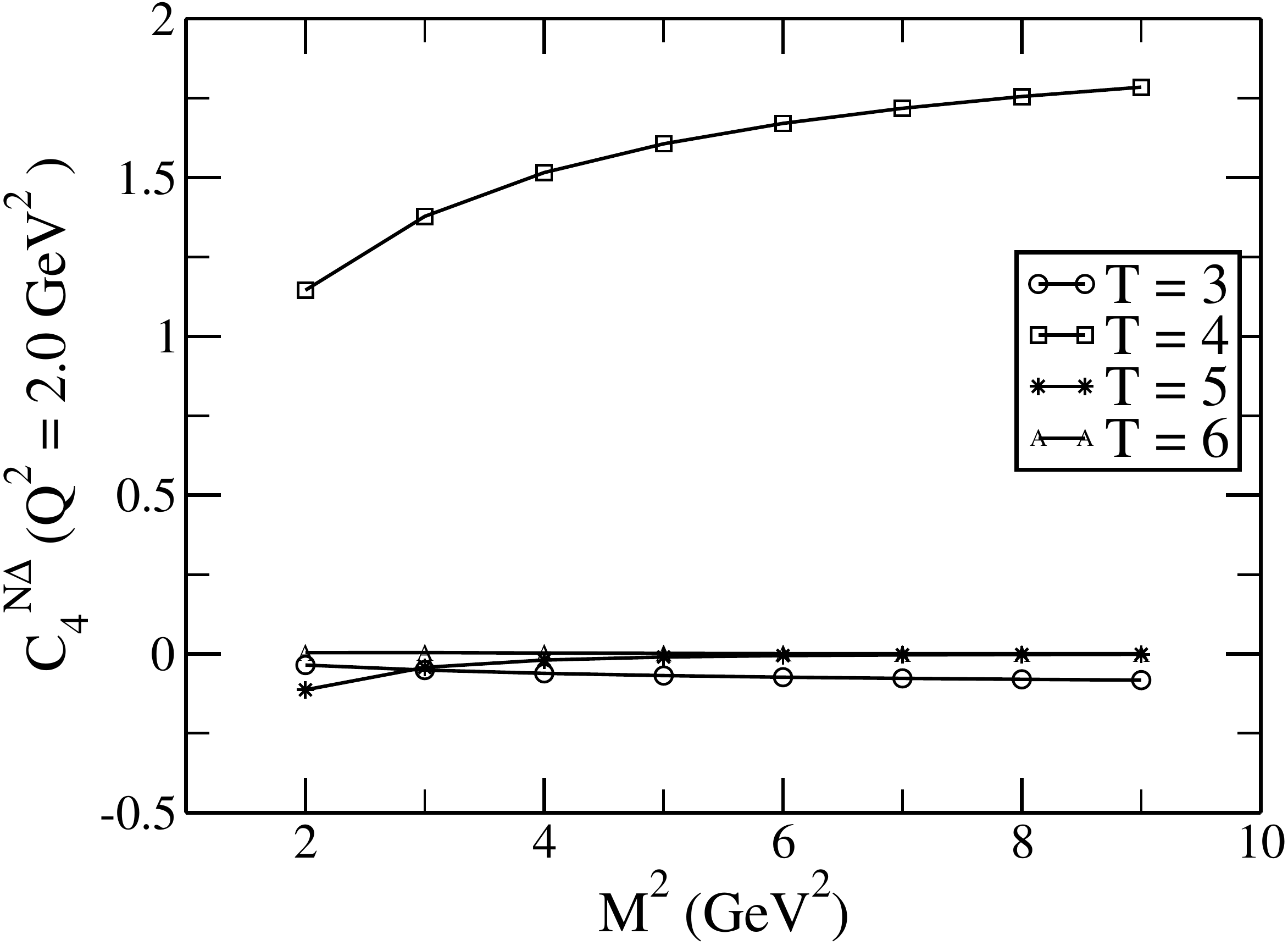}}
  \\
   \subfloat[]{\label{fig:Nc5con.eps}\includegraphics[width=0.4\textwidth]{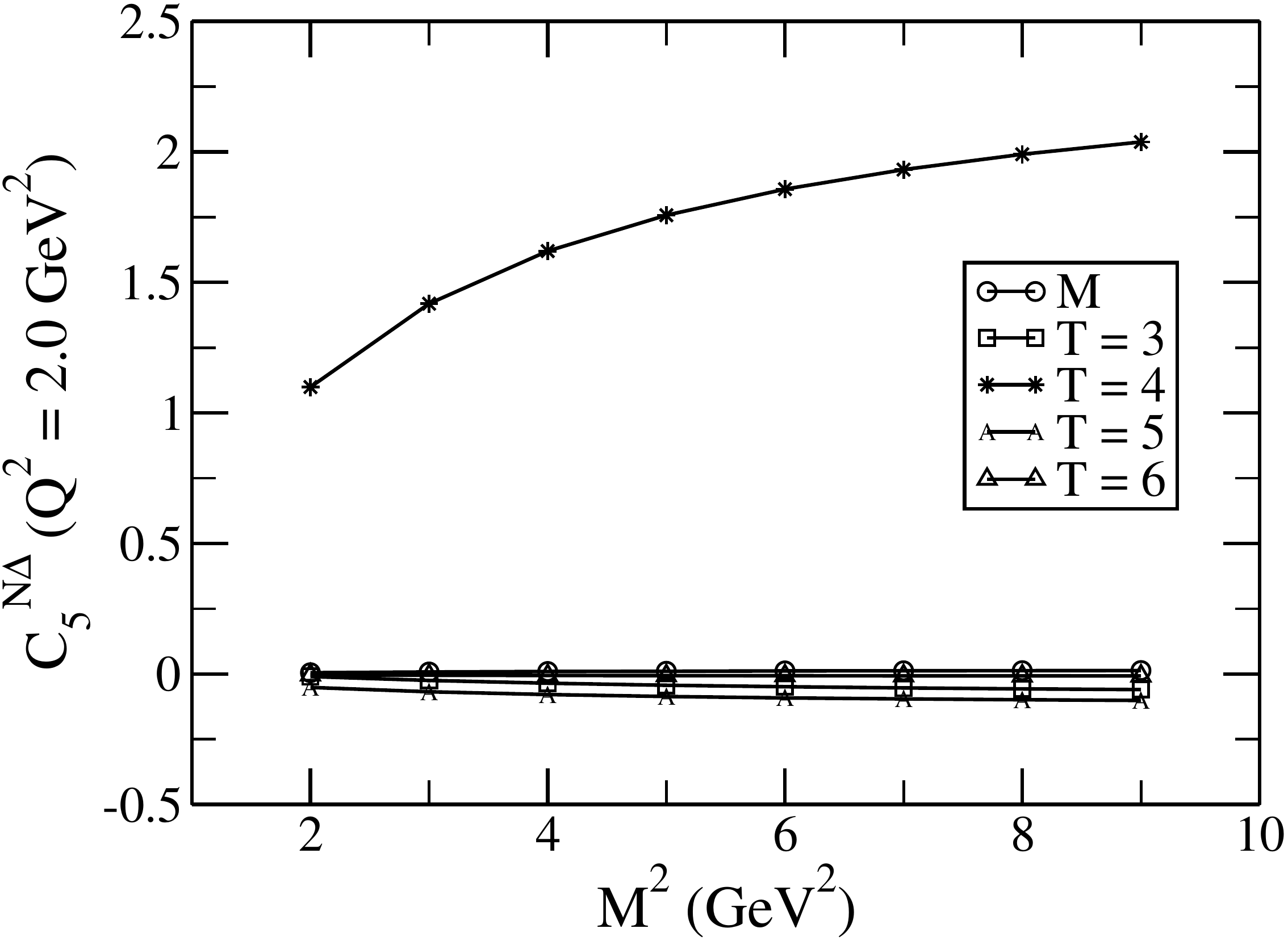}}
    \subfloat[]{\label{fig:Nc6con.eps}\includegraphics[width=0.4\textwidth]{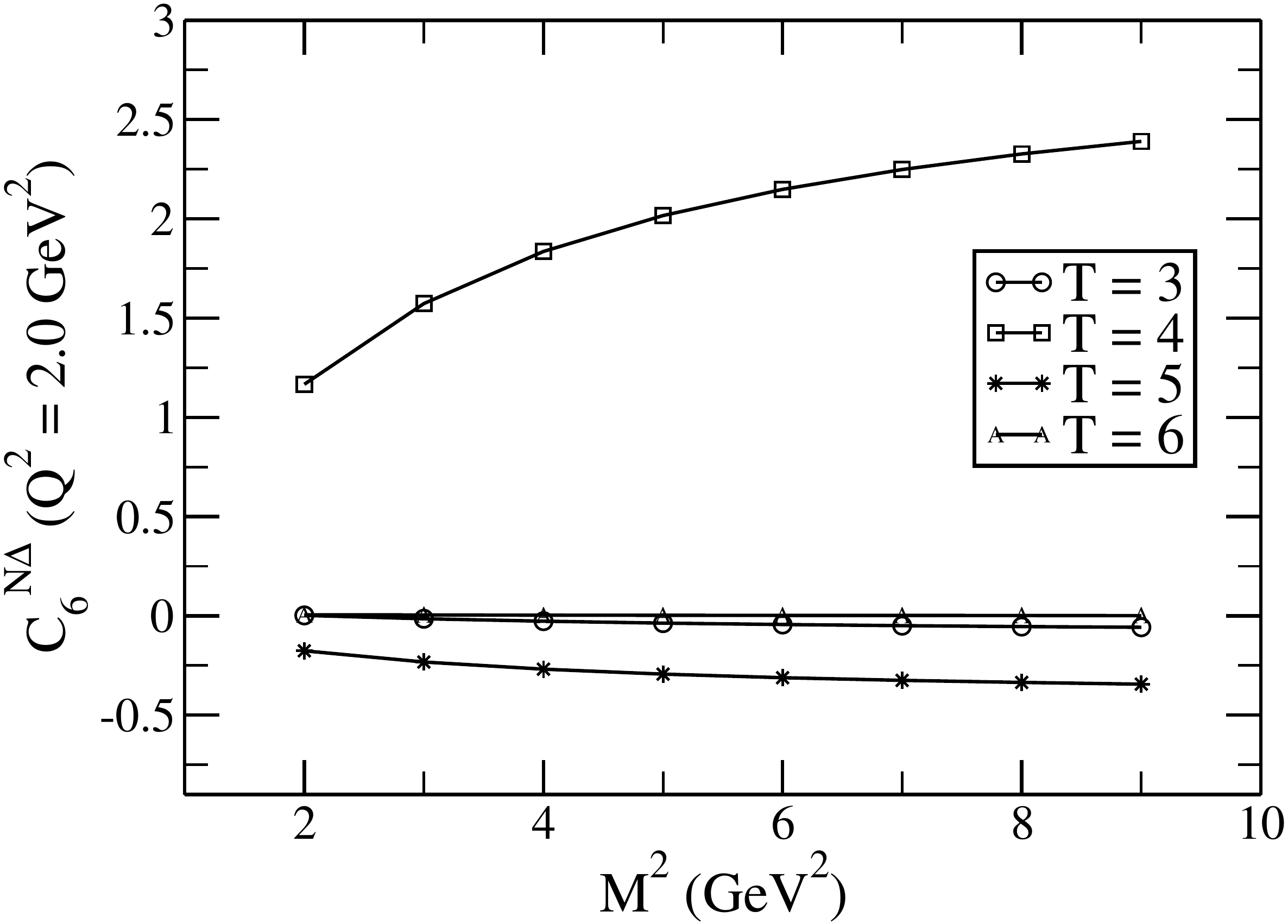}}
\caption{ The contribution to the  form factors at $Q^2 = 2.0~GeV^2$ from various twists.
(a) for $C_3^{N \Delta}$ form factors,
(b) for $C_4^{N \Delta}$ form factors,
(c) for $C_5^{N \Delta}$ form factors,
(d) for $C_6^{N \Delta}$ form factors.
The T3, T4, T5, T6 and M are twist-3, twist-4, twist-5, twist-6 and mass correction contributions, respectively.}
\label{fig:convergence}
\end{figure}

To analyse the convergence of the twist expansion, in Fig. (\ref{fig:convergence}), the contribution from definite twists to the form factors is presented at fixed $Q^2=2.0~GeV^2$ and $s_0=2.5~GeV^2$ values. As can be seen from the figures,
the dominant contribution to the form factors is from twist-4 DAs. Form factors $C_4^{N \Delta}$, $C_5^{N \Delta}$ and $C_6^{N \Delta}$ receive negligible contributions from other DAs. In the case of the $C_3^{N \Delta}$ form factors, the contribution of all the twists are comparable, and hence, one can not talk about the convergence of the twist expansion. Note that, for the case of $C_3^{N\Delta}$, although the twist expansion does not seem to converge, the mass predicted by the sum rules for the $C_3^{N\Delta}$ form factor is quite accurate as can be seen from Fig. \ref{fig:c3MDsq.pdf}. Indeed, the prediction of $M_\Delta$ using the sum rules for $C_3^{N\Delta}$ is more accurate than the predictions obtained from the sum rules for $C_4^{N\Delta}$ and $C_5^{N\Delta}$. This can be interpreted as an indication in favor of the reliability of the predictions on the form factors for $C_3^{N\Delta}$.
The baryon mass corrections to the DAs contribute only to the form factors $C_3^{N \Delta}$ and $C_5^{N \Delta}$.
The contribution of these correction to $C_5^{N \Delta}$ is negligible, whereas, the contribution to $C_3^{N \Delta}$ is twice the final result.
This means that the baryon mass correction contribution to this form factors even changes the sign of the form factor
and hence can not be neglected.

\begin{figure}[htp]
\centering
 \subfloat[]{\label{fig:NC3Mont.eps}
 \includegraphics[width=0.4\textwidth]{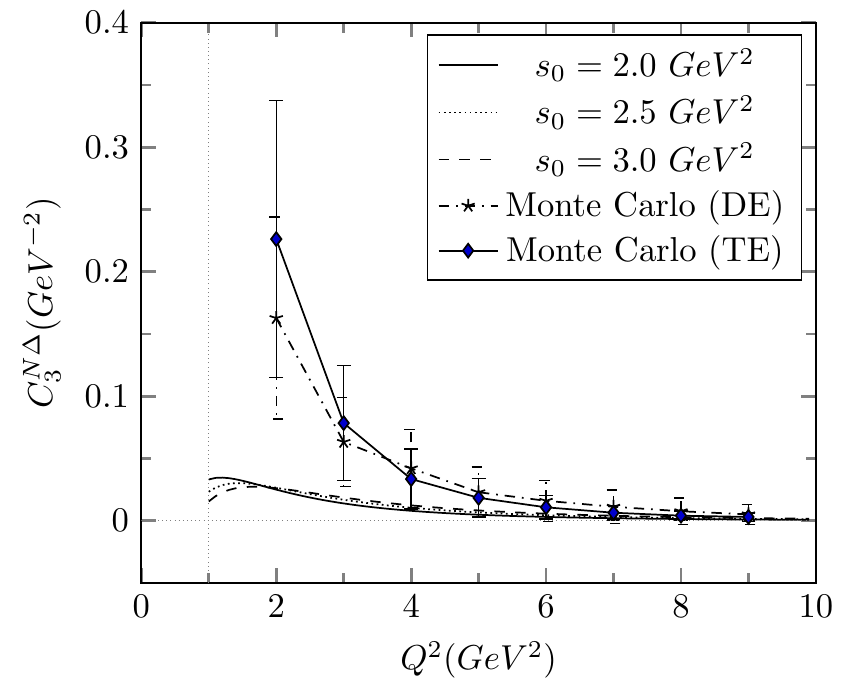}
 }
  \subfloat[]{\label{fig:NC4Mont.eps}
  \includegraphics[width=0.4\textwidth]{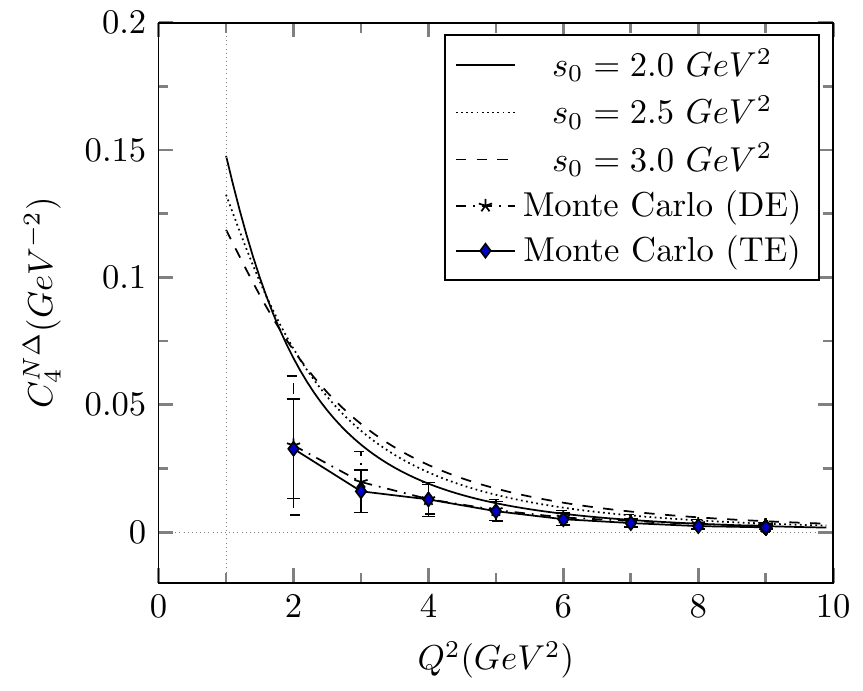}
}
  \\
   \subfloat[]{\label{fig:NC5Mont.eps}
   \includegraphics[width=0.4\textwidth]{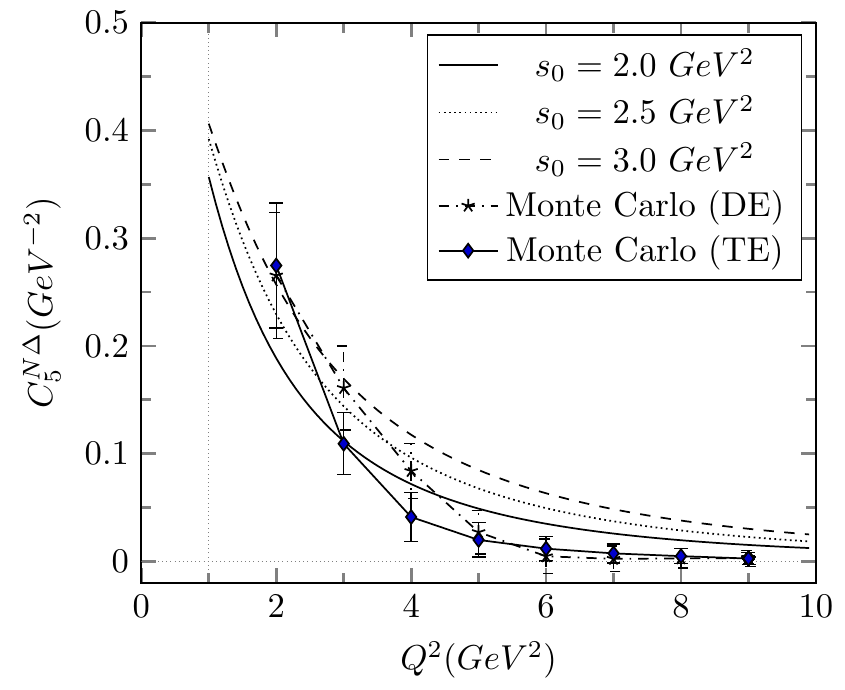}
   }
    \subfloat[]{\label{fig:NC6Mont.eps}
    \includegraphics[width=0.4\textwidth]{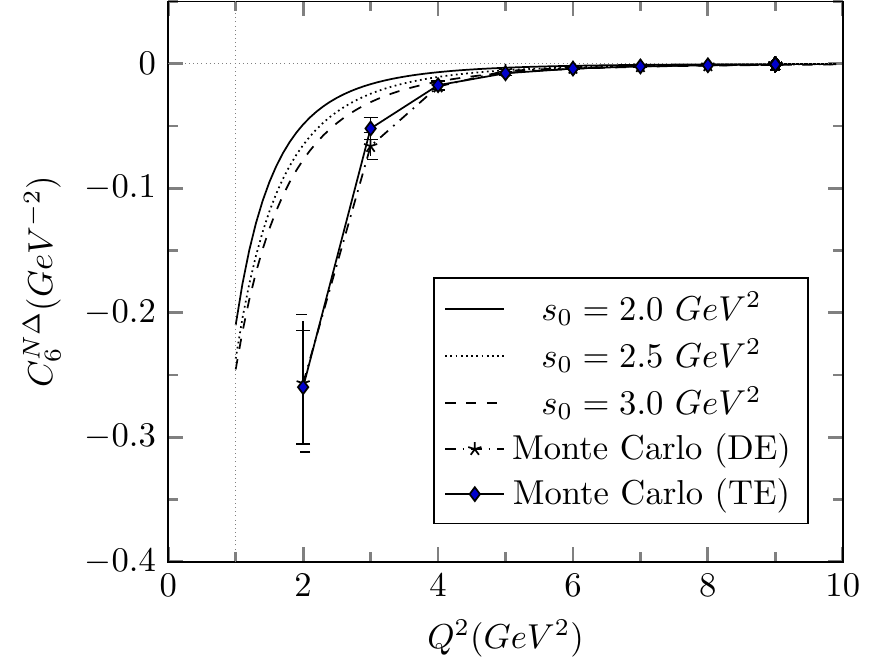}
}
\caption{The dependence of the form factors on $Q^2$. The results of the Monte Carlo  analysis are shown by symbols and error bars. Results of conventional analysis (calculated at $M^2=3.0~GeV^2$) are shown by lines.
(a) for $C_3^{N \Delta}$ form factor,
(b) for $C_4^{N \Delta}$ form factor,
(c) for $C_5^{N \Delta}$ form factor and
(d) for $C_6^{N \Delta}$ form factor.}
\label{fig:qsq}
\end{figure}

In Fig. (\ref{fig:qsq}), we present the $Q^2$ dependence of the form factors obtained
using two different types of analysis.
The results of the conventional  sum rules analysis is presented with lines for the central values of the parameters appearing in Table \ref{parameter_table}.
The circles and squares with error bars are the results of the Monte Carlo analysis.
The results obtained by both a double exponential (DE) and a triple exponential (TE) model of the correlation function is presented.
It is observed that for the form factors $C_3^{N \Delta}(Q^2)$, $C_4^{N \Delta}(Q^2)$ and $C_6^{N \Delta}(Q^2)$,
Monte Carlo analysis and the prediction for the conventional sum rules analysis agree with each other at large values of $Q^2$, but deviate from each other for small values of $Q^2$.
In the case of $C_5^{N \Delta}(Q^2)$, we observe that predictions for the central values agree at small $Q^2$,
but deviate from each other for large values of $Q^2$. Also, it is seen that although with the conventional sum rules analysis, one predicts  value for
form factor $C_3^{N \Delta}(Q^2)$ that is very close to zero, the Monte Carlo analysis shows that this form factor is consistent with significant non-zero values.
In the case of $C_4^{N \Delta}(Q^2)$, although the conventional sum rules analysis leads to a value that is significantly away from zero,
the Monte Carlo analysis shows that this form factor is almost consistent with zero.

The values of the form factors at zero momentum transfer $Q^2=0$ defines the corresponding charges. The sum rules method is only reliable at large enough values of $Q^2$, which is typically assumed to be $Q^2 > 1~GeV^2$. To obtain the value  at $Q^2=0$, the predictions has to be extrapolated.

The function that is typically used in the literature for the extrapolation has the form (see e.g. \cite{Hernandez:2010bx})
\begin{equation}
	C_i^{N \Delta}(Q^2) = \frac{C_i^{N \Delta}(0)}{(1+Q^2/m_{A}^2)^n}
	\label{eq:fit}
\end{equation}
where $n=2$. The axial mass $m_A$ is the free parameter which has been found experimentally to be \cite{Kitagaki:1990vs}
\begin {equation}
m_A = 1.28 _{-0.10}^{+0.08}.
\end{equation}
This value is used to fit our predictions to obtain $C_i^{N\Delta}$, $i=3,~4,~5$ or $6$. The obtained predictions on the axial charges are presented in Table \ref{fit_table}.

Although this functions gives a reasonable fit to our predictions on $C_3^{N\Delta}$, $C_4^{N\Delta}$ and $C_5^{N\Delta}$ using the conventional analysis, it fits poorly to the form factor  $C_6^{N\Delta}$. For this form factor,  a better fit is obtained for $n=4$. 
Furthermore,   close to $Q^2=1~GeV^2$, the predictions of the form factor from sum rules deviates from a pole form significantly. This can be interpreted as a failure of the sum rules at such low values of $Q^2$. For this reason, we fit the form factors only in the region $Q^2>2~GeV^2$. 


\begin{table}[t]
\begin{center}
\begin{tabular}{|c|c|c|c|}
\hline \hline
Form Factor & \begin{tabular}{c} $C_i^{N \Delta}(0) (GeV^{-2})$ \\ (Conventional Analysis) \end{tabular} & \begin{tabular}{c} $C_i^{N \Delta}(0) (GeV^{-2})$ \\ (MCDE) \end{tabular} & \begin{tabular}{c} $C_i^{N \Delta}(0) (GeV^{-2})$ \\ (MCTE) \end{tabular}   \\
\hline\hline
$C_3$ &	$0.11 \pm 0.03$ & $0.41\pm0.12$ & $0.22\pm0.07$ 	\\
\hline
$C_4$ &$0.27 \pm 0.09$	  &  $0.13\pm0.02$&  $0.10\pm0.02$\\
\hline
$C_5$ & $1.14\pm 0.20$  & $0.59\pm0.11$& $0.45\pm0.09$ \\
\hline
$C_6$ & $-1.65 \pm 0.46$  &$-1.88\pm0.15$ &  $-1.96\pm0.16$\\
\hline\hline
\end{tabular}
\caption{The values of dipole fit parameters, $C_i^{N \Delta}(0)$  for axial  form factor obtained from the conventional analysis, MC analysis using DE (MCDE) and MC analysis using TE (MCTE).}
	\label{fit_table}
\end{center}
\end{table}

At $Q^2=0$, experimentally the most easily accessible form factor is  $C_5^{N \Delta}$ \cite{Hernandez:2010bx}. This form factor also determines the axial charge of the transition $N-\Delta$.
Various predictions for this axial charge are as follows:
the off-diagonal Goldberger-Treiman relation predicts that the axial charge is $C_5^{N\Delta}(0) \simeq 1.20$,
the prediction of quark model ranges from  $C_5^{N\Delta}(0)$ = $0.81$ to $1.53$ (see e.g. \cite{BarquillaCano:2007yk} and reference therein),
in the case of chiral perturbation theory it is estimated to be $C_5^{N\Delta}(0)$ = $1.16$ \cite{Geng:2008bm},
 lattice QCD predicts $C_5^{N\Delta}(0)$ = $0.9 \pm 0.02$ \cite{Alexandrou:2006mc} and
the results from weak pion production is $C_5^{N\Delta}(0)$ = $1.08 \pm 0.1$ \cite{Hernandez:2010bx}
and $C_5^{N\Delta}(0)$ = $1.19 \pm 0.08$ \cite{Graczyk:2009qm}. Although these values are consistent with our prediction on $C_5^{N\Delta}(0)$ using the conventional analysis within error bars, the Monte Carlo analysis predicts smaller values. 
We can also compare our results Adler's model predictions \cite{Adler:1968tw}. There, the axial form factors  for the N $\rightarrow \Delta$ transition have been parameterized as
\begin{equation} 
C_j ^A (Q^2)= \frac{C_j^A(0)(1-a_jQ^2/(b_f-Q^2))}{(1-Q^2/m_A^2)^2};   ~~~~ j = 3, 4, 5
\end{equation}
with\\

$C_3^{N\Delta}(0) = 0$, $ C_4^{N\Delta}(0) = -\frac{C_5^{N\Delta}(0)}{4}= -0.3$ and $C_5^{N\Delta}(0) = 1.2$\\

$a_3 = b_3 = 0$, $a_4 = a_5 = -1.21$, $b_4 = b_5 = 2~ GeV^2$.\\

We see the predictions of the conventional analysis are consistent with the predictions of Adler's model except of the $C_3^{N\Delta}$, however, the predictions of the Monte Carlo methods are not consistent. 


In \cite{Androic:2012doa}, $G_A^{N \Delta}$, defined as:
\begin{eqnarray}
G_A^{N\Delta}(Q^2) &= \frac12 \left[M_N^2 - M_\Delta^2+Q^2\right] C_4^A(Q^2) - M_N^2 C_5^A(Q^2),
\end{eqnarray}
is measured to be $G_A^{N\Delta}(Q^2=0.34~GeV^2)= -  0.05\pm(0.35)_{stat} \pm(0.34)_{sys} \pm(0.06)_{theory}$. Using the values in Table \ref{fit_table}, $G_A^{N\Delta}=-1.04 \pm 0.19$ , $G_A^{N\Delta}=-0.54 \pm 0.10$ , $G_A^{N\Delta}=-0.41 \pm 0.08$ respectively using conventional analysis, MC analysis with DE, and MC analysis with TE. 
Note that, the experimental measurement has large error bars, but is consistent with $G_A^{N\Delta} =0$. The predictions of both the conventional analysis and MC analysis are consistent with the  experimental value of $G_A^{N\Delta}$ (within error bars), but the MC analysis predicts a much smaller value for $G_A^{N\Delta}$.

In conclusion, we have extracted the isovector axial-vector form factors of N-$\Delta$ transition by applying the LCSR.
The $Q^2$ dependence of form factors are obtained using the conventional analysis, and Monte Carlo analysis. The Monte Carlo analysis showed that the error bars in the predictions of the form factors are large especially in the low $Q^2$ region. This especially implies that the extrapolation to $Q^2=0$ is unreliable. On the other hand, it is shown that the predictions obtained by the conventional analysis is consistent with the results in the literature, when they exist. 

\section*{Acknowledgments}
 The work of U.O. is supported by the Scientific and Technological Research
Council of Turkey (TUBITAK) under the program 2214/A. Also, U.O. would like to thank the JINR, for warm hospitality.



\appendix
\newpage
\section{Explicit forms of the functions $F_i$ for the $N \rightarrow \Delta$ transition}
		         \begin{align*}
		         \begin{split}
		        &F_1=\int_0^{\beta} d\alpha\int_\alpha^{1}dx_2\int_0^{1-x_2}dx_1 2 [T_1-T_3-T_4+T_6-T_7-T_8-A_1+A_2-A_3-A_4+A_5-A_6]\\
		             & \qquad(x_1,x_2,1-x_1-x_2),\\
			&F_2=\int_\alpha^{1}dx_2\int_o^{1-x_2}dx_1  4 [A_1-A_2+A_3-T_1+T_3+T_7](x_1,x_2,1-x_1-x_2),\\
			&F_3=\int_o^{\beta}d\alpha\int_\alpha^{1}dx_3\int_o^{1-x_3}dx_1 2\left[V_1-V_2-V_3-V_4-V_5+V_6+T_1-T_3-T_4
			+T_6-T_7-T_8\right.\\    &\qquad+A_1-A_2+A_3+A_4-A_5+A_6](x_1,1-x_1-x_3,x_3),\\
			&F_4=\int_\alpha^{1}dx_3\int_0^{1-x_3}dx_1  2[V_1-V_2-V_3+A_1-A_2+A_3+T_1-T_3-T_7](x_1,1-x_1-x_3,x_3),\\
			&F_5=\int_o^{\beta}d\alpha\int_\alpha^{1}dx_3\int_o^{1-x_3} dx_12[T_1-T_3-T_4+T_6-T_7-T_8-A_1+A_2-A_3-A_4+A_5-A_6]\\
			    &\qquad(x_1,1-x_1-x_3,x_3),\\
			&F_6= \int_\alpha^{1}dx_3\int_\alpha^{1-x_3}dx_1   2[T_1-T_3-T_7](x_1,1-x_1-x_3,x_3),\\
			&F_7=\int_o^{\beta}d\alpha\int_\alpha^{1}dx_2\int_o^{1-x_2}dx_1 2[-A_1+A_2-A_3-A_4+A_5-A_6+T_1-T_3-T_4+T_6-T_7- T_8]\\
			   & \qquad(x_1,x_2,1-x_1-x_2),\\
			&F_8=\int_\alpha^{1}dx_2\int_\alpha^{1-x_2}dx_1[2V_1-2V_2-2V_3+2A_1-2A_2+2A_3-4T_1+4T_3+4T_7](x_1,x_2,1-x_1-x_2),\\  	
			&F_9=\int_\alpha^1 dx_2\int_{0}^{1-x_2}dx_1[2T_1-T_3-T_4-T_7-T_8+A_3-A_4-S_1+S_2+P_1 -P_2-2V_1+2V_2+V_3+V_4]\\
                        &\qquad(x_1,x_2,1-x_1-x_2),\\
			&F_{10}= \int_{\alpha}^{1}dx_3\int_{0}^{1-x_3}dx_1[-V_1+V_3+V_5+T_1-T_3-T_7-1/2~A_1-1/2~A_3-1/2~A_5](x_1,1-x_1-x_3,x_3),\\
			&F_{11}= \int_{0}^{1-x_2}dx_1[V_1 - T_1 ](x_1,x_2,1-x_1-x_2),\\
			&F_{12}= \int_{0}^{1-x_3}dx_1[V_1 - T_1+1/2~A_1 ](x_1,1-x_1-x_3,x_3),\\
			&F_{13}= \int_{0}^{1-x_2}dx_1[V_1^{M}-T_1^{M}](x_1,x_2,1-x_1-x_2),\\
			&F_{14}= \int_{0}^{1-x_3}dx_1[V_1^{M}-T_1^{M}+1/2~A_1^{M}](x_1,1-x_1-x_3,x_3),\\
			&F_{15}=\int_{0}^{\beta}d{\alpha}\int_{\alpha}^{1}dx_2\int_{0}^{1-x_2}dx_1[T_2-T_3-T_4+T_5+T_7+T_8](x_1,x_2,1-x_1-x_2),\\
                        &F_{16}=\int_{0}^{\beta}d{\alpha}\int_{\alpha}^{1}dx_3\int_{0}^{1-x_3}dx_1[-T_2+T_3+T_4-T_5-T_7-T_8](x_1,1-x_1-x_3,x_3),\\
			\end{split}
			\end{align*}			
			\newpage
			\begin{align*}
&F_{17}=\int_{0}^{\beta}d{\alpha}\int_{\alpha}^{1}dx_2\int_{0}^{1-x_2}dx_1[-T_1-T_2+2T_3+2T_4-T_5-T_6+2A_1-2A_2-2A_5+2A_6]\\
&\qquad(x_1,x_2,1-x_1-x_2),\\
			&F_{18}=\int_{0}^{\beta}d{\alpha}\int_{\alpha}^{1}dx_3\int_{0}^{1-x_3}dx_1[T_1-T_2-T_5+T_6-2T_7-2T_8](x_1,1-x_1-x_3,x_3),\\
			&F_{19}=\int_{0}^{1-x_2}d{x_1}[V_1^{M}-T_1^{M}](x_1,x_2,1-x_1-x_2),\\
&F_{20}=\int_{0}^{1-x_3}d{x_1}[V_1^{M}](x_1,1-x_1-x_2,x_3),\\
			&F_{21}=\int_0^{1-x_2}d{x_1}[P_1-S_1+V_1+V_2-A_1+A_2-T_3-T_7](x_1,x_2,1-x_1-x_2),\\		
		         &F_{22}=\int_0^{1-x_3}d{x_1}\left[V_3-T_1-A_3\right](x_1,1-x_1-x_3,x_3),\\ &F_{23}=\int_{0}^{\beta}d{\alpha}\int_{\alpha}^{1}dx_2\int_{0}^{1-x_2}dx_1[-T_1+T_3+T_4-T_6+T_7+T_8+A_1-A_2+A_3+A_4-A_5+A_6]\\
&\qquad(x_1,x_2,1-x_1-x_2),\\
			&F_{24}=\int_{0}^{\beta}d{\alpha}\int_{\alpha}^{1}dx_3\int_{0}^{1-x_3}dx_1[T_1-T_3-T_4+T_6-T_7-T_8\\
			&\qquad+A_1-A_2+A_3+A_4-A_5+A_6+V_1-V_2-V_3-V_4-V_5+V_6](x_1,1-x_1-x_3,x_3),\\
			&F_{25}= \int_{0}^{1-x_2}dx_1[S_1-P_1-V_3-2V_2-A_3+T_3+T_7(x_1,x_2,1-x_1-x_2),\\
			&F_{26}= \int_{0}^{1-x_3}dx_1[-V_3 + T_1+T_7+A_3](x_1,1-x_1-x_3,x_3),\\
			&F_{27}=\int_{0}^{1-x_2}d{x_1}[-2V_1^{M}+T_1^{M}](x_1,x_2,1-x_1-x_2),\\
			&F_{28}=\int_{0}^{1-x_3}d{x_1}[T_1^{M}](x_1,1-x_1-x_2,x_3),\\
			&F_{29}= \int_{0}^{\beta}d{\alpha}\int_{\alpha}^{1}dx_2\int_{0}^{1-x_2}dx_1[-T_1+T_2+T_5-T_6+2T_7+2T_8](x_1,x_2,1-x_1-x_2),\\
			&F_{30}=\int_{0}^{\beta}d{\alpha}\int_{\alpha}^{1}dx_3\int_{0}^{1-x_3}dx_1[T_2-T_3-T_4+T_5+T_7+T_8-V_1+V_2+V_3
			+V_4+V_5-V_6\\
			&\qquad -A_1+A_2+A_3-A_4+A_5-A_6](x_1,1-x_1-x_3,x_3),\\			
			&F_{31}=\int_\alpha^{1}dx_2\int_\alpha^{1-x_2}dx_1[2T_1-T_4-T_7-T_8+A_3-A_4-S_1+S_2\\			&\qquad+P_1-P_2-2V_1+2V_2+V_3+V_4](x_1,x_2,1-x_1-x_2),\\
			&F_{32}= \int_\alpha^{1}dx_3\int_\alpha^{1-x_3}dx_1[T_1-T_3-T_7-V_1+V_3+V_5-A_1-A_3-A_5](x_1,1-x_1-x_3,x_3),\\ 	
			\end{align*}
\newpage
\begin{align*}
&F_{33}=\int_0^{1-x_2}d{x_1}[V_1-T_1](x_1,x_2,1-x_1-x_2),\\
&F_{34}=\int_0^{1-x_3}d{x_1}[V_1-T_1+A_1](x_1,1-x_1-x_3,x_3),\\
			&F_{35}=\int_{0}^{1-x_2}d{x_1}[V_1^{M}-T_1^{M}](x_1,x_2,1-x_1-x_2),\\
			&F_{36}=\int_{0}^{1-x_3}d{x_1}[V_1^{M}-T_1^{M}+A_1^{M}](x_1,1-x_1-x_2,x_3),\\
			&F_{37}= \int_{0}^{\beta}d{\alpha}\int_{\alpha}^{1}dx_2\int_{0}^{1-x_2}dx_1[T_2-T_3-T_4+T_5+T_7+T_8](x_1,x_2,1-x_1-x_2),\\
			&F_{38}=\int_{0}^{\beta}d{\alpha}\int_{\alpha}^{1}dx_3\int_{0}^{1-x_3}dx_1[T_2-T_3-T_4+T_5+T_7+T_8](x_1,1-x_1-x_3,x_3),\\
			&F_{39}= \int_{0}^{\beta}d{\alpha}\int_{\alpha}^{1}dx_2\int_{0}^{1-x_2}dx_1[T_1-T_3-T_4+T_6-T_7-T_8-A_1+A_2\\
			&\qquad -A_3-A_4+A_5-A_6](x_1,x_2,1-x_1-x_2),\\
&F_{40}=\int_{0}^{\beta}d{\alpha}\int_{\alpha}^{1}dx_3\int_{0}^{1-x_3}dx_1[T_1-T_3-T_4+T_6-T_7-T_8+A_1-A_2+A_3\\
			&\qquad+A_4-A_5+A_6-V_1+V_2+V_3+V_4+V_5-V_6](x_1,1-x_1-x_3,x_3),\\
			&F_{41}=\int_\alpha^{1}dx_2\int_\alpha^{1-x_2}dx_1[A_1-A_2+A_3-V_1+V_2+V_3](x_1,x_2,1-x_1-x_2),\\
			&F_{42}= \int_\alpha^{1-x_3}dx_1 [T_1-T_3+T_7](x_1,1-x_1-x_3,x_3),\\
\end{align*}
			\end{document}